\documentclass[preprint, a4paper, times]{aastex61}
\usepackage{natbib}
\newcommand{\betaloss}{\beta_{\mathrm{loss}}}
\newcommand{\betaprod}{\beta_{\mathrm{prod}}}
\newcommand{\urel}{u_{\mathrm{rel}}}
\newcommand{\sigCX}{\sigma\!_{\mathrm{cx}}}
\newcommand{\VA}{V\!_{\mathrm{A}}}

\newcommand{\kms}{km~s$^{-1}$}
\newcommand{\hepl}{He$^+$\,}
\newcommand{\nHe}{n_{\mathrm{He}}}
\newcommand{\nHepl}{n_{\mathrm{He}^+}}
\newcommand{\uTHepl}{u_{\mathrm{T,He}^+}}
\newcommand{\uTHe}{u_{\mathrm{T,He}}}

\received{July 1, 2016}
\revised{September 27, 2016}
\accepted{\today}
\submitjournal{ApJ}
\begin{document}

\title{The helium Warm Breeze in IBEX observations as a result of
charge exchange collisions in the outer heliosheath} 

\correspondingauthor{Maciej Bzowski}
\email{bzowski@cbk.waw.pl}

\author{Maciej Bzowski}
\affiliation{Space Research Centre PAS (CBK PAN) \\
Bartycka 18A \\
00-716 Warsaw, Poland}

\author{Marzena A. Kubiak}
\affiliation{Space Research Centre PAS (CBK PAN) \\
Bartycka 18A \\
00-716 Warsaw, Poland}

\author{Andrzej Czechowski}
\affiliation{Space Research Centre PAS (CBK PAN) \\
Bartycka 18A \\
00-716 Warsaw, Poland}

\author{Jolanta Grygorczuk}
\affiliation{Space Research Centre PAS (CBK PAN) \\
Bartycka 18A \\
00-716 Warsaw, Poland}

\begin{abstract} 

We simulated the signal due to neutral He atoms, observed by 
Interstellar Boundary Explorer (IBEX), assuming that charge exchange collisions 
between neutral He atoms and \hepl ions operate everywhere between the heliopause 
and a distant source region in the local interstellar cloud, where the neutral and 
charged components are in thermal equilibrium. We simulated several test cases of 
the plasma flow within the outer heliosheath and investigated the signal 
generation for plasma flows both in the absence and in the presence of the 
interstellar magnetic field. We found that a signal in the portion of IBEX data 
identified as due to the Warm Breeze does not arise when a homogeneous plasma flow 
in front of the heliopause is assumed, but it appears immediately when any 
reasonable disturbance in its flow due to the presence of the heliosphere is 
assumed. We obtained a good qualitative agreement between the data selected for 
comparison and the simulations for a model flow with the velocity vector of the 
unperturbed gas and the direction and intensity of magnetic field adopted from 
recent determinations. We conclude that direct-sampling observations of neutral He 
atoms at 1~AU from the Sun are a sensitive tool for investigating the flow of interstellar matter in the outer heliosheath, that the Warm Breeze is indeed the secondary 
population of interstellar helium, as it was hypothesized earlier, and that the WB signal is consistent with the heliosphere distorted from axial symmetry by the interstellar magnetic field.

\end{abstract}

\keywords{will be inserted later}

\section{Introduction} 
\label{sec:intro}

The Warm Breeze (WB) is a population of neutral He atoms that are flowing into the 
heliosphere in addition to the well-known population of pristine interstellar 
neutral (ISN) gas \citep{fahr:68,bertaux_blamont:71}. The existence of WB was 
discovered by \citet{bzowski_etal:12a} based on observations carried out using the 
IBEX-Lo neutral-atom detector \citep{fuselier_etal:09b} on board of the Interstellar 
Boundary Explorer (IBEX) space probe \citep{mccomas_etal:09a}. A 
reconnaissance study of the WB properties was presented by 
\citet{kubiak_etal:14a}. These latter authors suggested that WB is twice slower 
and twice warmer than the pristine ISN gas, it is coming from a direction 
different by $\sim 20 \degr$ to that of the ISN gas, and its density in front of 
the heliosphere is a few percent of that of ISN He. Both \citet{bzowski_etal:12a} 
and \citet{kubiak_etal:14a} considered a hypothesis that WB is the secondary 
population of interstellar neutral gas, similar to the secondary population of 
interstellar hydrogen, created in the outer heliosheath due to charge-exchange 
collisions between the unperturbed neutral population of interstellar matter and 
the heated, compressed, and deflected population of interstellar plasma in the 
outer heliosheath (OHS).

The existence of the secondary population of interstellar gas was suggested 
by \citet{baranov_malama:93} for ISN H. This population, discovered indirectly in 
observations of the heliospheric backscatter glow in the Lyman-$\alpha$ line 
\citep{lallement_bertaux:90a, lallement_etal:93a}, is produced in copious amounts 
in OHS due to the relatively high densities of the reaction substrates and the 
large reaction cross section \citep{lindsay_stebbings:05a}. The existence of a 
helium equivalent for the hydrogen secondary population can be expected as a 
result of several candidate charge exchange reactions \citep[see Figure 24 and 
Table 2 in ][]{bzowski_etal:12a}. The most obvious candidate seemed to be: 
\begin{equation} \mathrm{H_{OHS}} + \mathrm{He^+_{OHS}} \rightarrow 
\mathrm{He_{sec}} + \mathrm{H^+_{OHS}}, \label{eq:cxH-HePl} \end{equation} and it 
was considered by \citet{mueller_zank:03a, mueller_zank:04a}, who pointed out that 
the density of the secondary helium population expected in the heliosphere due to 
this reaction should be very low: about 1\% of that of the primary ISN He. This 
was because of the very low reaction cross section for the collision speeds 
expected in the outer heliosheath \citep{barnett_etal:90}. However, 
\citet{bzowski_etal:12a} pointed out that the most effective source of the 
secondary He is expected to be the charge exchange reaction between the He$^+$ 
ions and He atoms: \begin{equation} \mathrm{He_{ISN}} + \mathrm{He^+_{OHS}} 
\rightarrow \mathrm{He_{sec}} + \mathrm{He^+_{OHS}} \label{eq:cxHe-Hepl} 
\end{equation} because even though the absolute densities of the He atoms and 
He$^+$ ions in the outer heliosheath are an order of magnitude lower than these of 
H and H$^+$, the reaction cross section is larger by a factor of 250.

\citet{kubiak_etal:16a} analyzed data from observations by IBEX-Lo performed from 
2009 to 2014 and determined the temperature, density and velocity vector of WB 
much more precisely than it was possible for \citet{kubiak_etal:14a}. In addition, 
at the time of the analysis by \citet{kubiak_etal:16a} results of precise 
determination of the inflow direction of the unperturbed primary ISN He became 
available \citep{bzowski_etal:14a, wood_etal:15a, bzowski_etal:15a, 
mobius_etal:15b, schwadron_etal:15a, mccomas_etal:15b}. Based on this insight, 
\citet{kubiak_etal:16a} found that the inflow directions of ISN He, ISN H 
\citet[available from]{lallement_etal:05a, lallement_etal:10a}, and WB are 
coplanar and that the plane in which these directions are included also includes 
the center of the IBEX Ribbon (\citet{mccomas_etal:09a}; see also 
\citet{funsten_etal:13a, funsten_etal:15a}). Later on, it was demonstrated that 
also the inflow direction of ISN O is within this plane \citep{schwadron_etal:16a} 
and that WB has its oxygen counterpart \citep{park_etal:16a}. All this is a very 
strong evidence in favor of the hypothesis that WB is indeed the secondary 
population of ISN He.

This is because most of the state of the art heliospheric models, which treat the 
heliospheric and interstellar neutral populations kinetically 
\citep[e.g.,][]{izmodenov_etal:05a, pogorelov_etal:08a}, suggest that the 
direction of inflow of the heliospheric secondary population should be deflected 
from that of the primary population in the plane defined by the vectors of Sun's 
motion through the local interstellar medium and of the local interstellar 
magnetic field (ISMF). As a result of the action of ISMF, the heliosphere should 
be distorted from axial symmetry \citep{ratkiewicz_etal:98a}, 
and the distribution of pressure, temperature, flow vectors of the plasma, and 
magnetic field intensity should be modified accordingly. The distortion predicted 
by purely-MHD models is mediated by the effects of charge-exchange between the 
plasma and ISN atoms \citep[see also ][]{izmodenov_alexashov:06a, 
izmodenov_alexashov:15a, pogorelov_etal:06b}.

The direction of ISMF is expected to conform with the IBEX Ribbon direction, if 
the Ribbon is created via the secondary ENA emission mechanism 
\citep{heerikhuisen_etal:10a}. This hypothesis is in agreement with in situ 
measurements of ISMF by Voyager \citep{burlaga_ness:14a, burlaga_ness:14b} based 
as well on simple geometrical arguments \citep{grygorczuk_etal:14a} as on 
analysis performed using sophisticated heliospheric models 
\citep{zirnstein_etal:16b}. In the secondary ENA emission mechanism, as suggested 
by \citet{mccomas_etal:09a} and \citet{schwadron_etal:09b}, the energetic neutral 
atoms (ENA) that are running radially away from the heliosphere are re-ionized 
beyond the heliopause and picked up by ISMF. They begin to gyrate around the field 
lines and eventually are re-neutralized due to charge exchange with atoms from the 
ambient ISN gas and run away in all directions, including those towards the Sun 
and the IBEX detectors. In the regions where the original ENA directions are close 
to perpendicular to the field lines, several scenarios 
\citep{heerikhuisen_etal:10a, mccomas_etal:09b, schwadron_etal:09b, 
chalov_etal:10a} have been proposed in which the signal observed by IBEX should be 
amplified with respect to the global ENA flux. Crucial in these scenarios is the 
geometry of the magnetic field. In all these scenarios, the source regions of the 
IBEX Ribbon is located close beyond the heliopause, its projection on the sky is 
close to circular, and the direction of ISMF is close to the center of this 
structure.

These scenarios, and especially that suggested by \citet{heerikhuisen_etal:10a}, 
gained credibility when \citet{swaczyna_etal:16a} determined parallax of the 
Ribbon and found that indeed, the Ribbon source is located at $\sim 140$~AU, i.e., 
beyond the distance to the heliopause found by Voyager 1 \citep{stone_etal:13a, 
gurnett_etal:13a}. Furthermore, \citet{swaczyna_etal:16b} were able to reproduce 
the dependence of the position of Ribbon center on ENA energy, discovered by 
\citet{funsten_etal:13a, funsten_etal:15a}. This was done using an analytic model 
adapted from \citet{mobius_etal:13a}, which assumes the secondary ENA emission 
mechanism due to \citet{heerikhuisen_etal:10a}, and adopting a spatial structure 
of the ENA emission from the supersonic solar wind resulting from 
observation-based solar wind structure, obtained by \citet{sokol_etal:15d}.

These insights suggest a consistent physical scenario for the creation of the 
heliospheric boundary region, distorted from axial symmetry by the interstellar 
magnetic field, which forms due to interaction of the magnetized interstellar 
plasma with the latitudinally-structured solar wind and is mediated on one hand by 
the ISN gas, and on the other hand by the derivative charged populations of pickup 
ions, created in different regions inside and outside of the heliosphere, and by 
ENAs \citep[for review, see ][]{zank:15a}. In this scenario, the plasma conditions 
in OHS, where the secondary ISN He is created, are strongly anisotropic, as 
illustrated by \citet{heerikhuisen_etal:14a} and \citet{izmodenov_alexashov:15a}, 
with relatively strong spatial gradients of the plasma velocity (both in magnitude 
and the flow direction), density, and temperature. Hence the local production rate 
of the secondary population must be strongly anisotropic, and since the neutral 
atoms move freely across the magnetic field lines, then effectively the 
distribution function of the secondary population of ISN He must be expected 
neither homogeneous, nor isotropic.  

However, all these complexities were neglected by \citet{kubiak_etal:16a} in their 
analysis of WB observations. These authors used a sophisticated method of data 
analysis and parameter fitting \citep{swaczyna_etal:15a} and a high-fidelity 
numerical scheme \citep{sokol_etal:15b} to reproduce the data, but the physical 
model of WB they employed was, in fact, the so-called hot model of ISN gas, originally 
proposed by several authors in 1970s \citep[e.g.][]{fahr:78, wu_judge:79a}. This model was  
modified to account for a variation in the ionization losses with time as proposed 
by \citet{rucinski_etal:03}, and taking into account the most recent insight into 
the variation of the helium ionization rate \citep{bzowski_etal:13b, 
sokol_bzowski:14a, sokol_etal:16a}. However, the hot model assumes that the 
distribution function in the source region is given by the 
Maxwell-Boltzmann function with the parameters (bulk velocity, temperature, 
density) isotropic and homogeneous in space. \citet{kubiak_etal:16a}, as well as 
\citet{kubiak_etal:14a} assumed that the distribution function of WB conforms to 
these prerequisites and that the source region is located at 150~AU from the Sun.

With these assumptions, using the sophisticated model of the measurement process 
executed by IBEX, they were able to successfully reproduce the observed signal and 
found the parameters of this model with surprisingly good precision \citep[see the 
uncertainty discussion and data vs model comparison in][]{kubiak_etal:16a}. 
While the $\chi^2$ value they obtained statistically speaking is significantly too 
high, it is clear from inspection of the fit residuals shown by these authors that 
a large portion of this excess can be attributed to some residual non-He 
populations still present in the data. Generally, however, the model fits to data 
from all five IBEX observation seasons very well.

The high quality of data reproduction by the simple model used by \citet{kubiak_etal:16a} is intriguing especially given one of the alternative hypotheses proposed by \citet{kubiak_etal:14a} to explain WB, i.e., the hypothesis of a neutral He flow from a hypothetical nearby cloud-cloud interface. A mechanism potentially responsible for such a flow was proposed by \citet{grzedzielski_etal:10b} for hydrogen. \citet{kubiak_etal:14a} pointed out that for He it is even more plausible because due to a lower cross section for elastic scattering and a higher mass, He atoms will have a much larger path for thermalization, so the hypothetical cloud-cloud boundary layer may be located significantly farther away from the Sun than that needed in the original scenario proposed by \citet{grzedzielski_etal:10b}. The distribution function of such a flow would likely be relatively closely approximated by a Maxwell-Boltzmann distribution with a temperature higher than that of the ambient unperturbed ISN He, density much lower, and a different velocity relative to the Sun. 

The objective of this paper is to verify the hypothesis that WB is indeed the 
secondary population of ISN He, created beyond the heliopause due to 
charge-exchange collisions between ISN He$^+$ ions and ISN He atoms. To that end, 
we modify the simulation approach used by \citet{kubiak_etal:16a} and we assume 
that the distribution function of ISN He is given by the isotropic 
Maxwell-Boltzmann function not at 150~AU, as adopted by \citet{bzowski_etal:15a} 
and \citet{kubiak_etal:16a}, but much farther: at 5000~AU. With this, we adopt 
several simplified models of the plasma flow past the heliopause that have some of 
the features predicted by state of the art heliospheric models, and we allow for 
gains and losses of He atoms due to charge exchange along the trajectories of the ISN He atoms 
that actually reach the IBEX-Lo detector. We simulate (in a simplified way) the 
signal observed by IBEX-Lo in these different OHS models for several selected subsets of IBEX observations.

We demonstrate that under these assumptions the WB signal indeed is created and 
that it is sensitive to various aspects of the plasma in OHS, 
including the size of the OHS, its distortion from axial symmetry, the plasma 
temperature etc. We discuss the implications and suggest next steps in researching 
the interactions within and physical state of the interstellar matter in the 
boundary region of the heliosphere.

\section{Model} 
\label{sec:Model}

\subsection{The idea of the simulation} 
\label{sec:SignalSimulation}

In this section we start with an outline of the idea of the simulation of the signal observed by IBEX-Lo used by \citet{bzowski_etal:15a} and \citet{kubiak_etal:16a} to simulate the ISN He and WB populations. Subsequently, we present the differences introduced to the model for the purpose of our study.

The signal simulation process used by these authors was presented in detail by 
\citet{sokol_etal:15b}. Effectively, the signal (which is proportional to time-, 
energy-, and collimator-averaged differential flux) was calculated as an integral 
of the statistical weights $\omega_{\mathrm{LOC}}$ associated with a given ballistic trajectory 
of He atoms, determined by the location in space $\vec{r}_{\mathrm{obs}}$ of the 
detector and the heliocentric velocity vector $\vec{v}_{\mathrm{obs}}$ of the atom 
at $\vec{r}_{\mathrm{obs}}$. The atom velocity in the source region is 
$\vec{v}_{\mathrm{src}}$. The statistical weight $\omega_{\mathrm{loc}}$ was 
calculated as a product of the probability 
$f_{\mathrm{src}}(\vec{v}_{\mathrm{src}}(\vec{r}_{\mathrm{obs}}, 
\vec{v}_{\mathrm{obs}}))$ that an atom at this trajectory exists in the source 
region and of the probability $w_{\mathrm{ion}}(\vec{r}_{\mathrm{obs}}, 
\vec{v}_{\mathrm{obs}}, r_{\mathrm{src}})$ that it is not ionized during its 
travel from the source region to the detection site at space 
$\vec{r}_{\mathrm{obs}}$: 
\begin{equation} 
\omega_{\mathrm{loc}}(\vec{v}_{\mathrm{obs}}) = 
f_{\mathrm{src}}(\vec{v}_{\mathrm{src}})\,w_{\mathrm{ion}}(\vec{r}_{\mathrm{obs}}, 
\vec{v}_{\mathrm{obs}}, r_{\mathrm{src}}), 
\label{eq:omegaLocDef1} 
\end{equation} 
where $r_{\mathrm{src}}$ is the heliocentric distance of the source region. The 
relations between $\vec{v}_{\mathrm{obs}}, \vec{r}_{\mathrm{obs}}$ on one hand and 
$\vec{v}_{\mathrm{src}}$ on the other hand were obtained from solutions of the 
hyperbolic Kepler equation, obtained individually for all atoms, and the 
probabilities $w_{\mathrm{ion}}$ were calculated numerically along the 
trajectories taking into account the evolution of the ionization rate during the 
time needed to travel from the source region to the observation locus. The source 
distance $r_{\mathrm{src}}$ was fixed at 150~AU \citep[for discussion of this 
choice, see][]{mccomas_etal:15b}, which is just outside the heliopause. 
Consequently, the assumption that there is virtually no production of He atoms 
between the source and detection loci was acceptable.

The local distribution function of ISN He at $\sim 1$ AU, where it is sampled by IBEX, is strongly anisotropic, as illustrated by \citet{mueller_etal:16a}, and therefore must be integrated numerically. To obtain the simulated signal, the statistical weights $\omega_{\mathrm{loc}}$ were subsequently integrated over atom speeds relative to the detector and over the collimator transmission function $T_{\mathrm{coll}}$ to obtain the signal characteristic for a given line of sight. Since IBEX is a spinning spacecraft in an elongated orbit around the Earth, with the spin axis directed towards the Sun, the boresight of the IBEX-Lo instrument is continuously scanning a great circle in the sky. The spin period is divided into a fixed number of time intervals, which effectively correspond to fixed bins in the spin 
angle, which correspond to fixed regions in the sky. The IBEX spin axis is moved 
every few days to follow the Sun (once or twice per IBEX orbit), so at each orbit 
the instrument observes a different portion of the sky. To simulate this, the 
speed- and collimator-averaged flux was subsequently averaged over a selected spin 
angle bin $\Delta \psi$. Since the observations adopted for analysis are taken 
during several intervals of ''good times'' $\Delta t_i$, the simulated signal 
$F_{\mathrm{sim}}$ was further averaged over these time intervals. Effectively, 
the simulated signal for a given bin and a given orbit was calculated as:
 \begin{equation} 
F_{\mathrm{sim}} = \frac{C \sum \limits_i \int \limits_{\Delta 
t_i} \int \limits_{\Delta \psi} \int\limits_{\Delta \Omega_{\mathrm coll}} \int 
\limits_{\Delta v_{\mathrm{rel}}}\!\! \! T_{\mathrm{coll}} \, v_{\mathrm{rel}}^3\, 
\omega_{\mathrm{loc}}\, dv_{\mathrm{rel}}\, d \Omega_{\mathrm{coll}} \, d \psi\, d 
t }{\sum \limits_i \Delta t_i }, 
\label{eq:defSimSig} 
\end{equation} 
where $C$ is 
a proportionality constant. All the necessary coordinate and reference system 
transformations are presented by \citet{sokol_etal:15b}, who also discuss at 
length the important role of performing all of the above-mentioned integrations in 
obtaining a high-fidelity reproduction of the observed signal. A fundamental 
assumption in these calculations was that the parent distribution function at 
$r_{\mathrm{src}}$ is given by an isotropic Maxwell-Boltzmann function 
$f_{\mathrm{MB}}$ co-moving with the ISN gas, which flows with the velocity 
$\vec{u}_{\mathrm{LIC}}$ relative to the Sun and has a temperature 
$T_{\mathrm{LIC}}$ and density $n_{\mathrm{He,LIC}}$, homogeneously distributed 
beyond the distance $r_{\mathrm{src}}$ from the Sun.

In the present approach, we set the source region of the observed atoms much 
farther away from the Sun, so far that the interstellar matter can be regarded as 
truly unaffected by any interaction with the solar output: $r_{\mathrm{src}} = 
5000$~AU. We follow a similar paradigm as outlined earlier, but now we allow for 
charge-exchange collisions to modify the statistical weights $\omega$ in the 
region outside the heliopause. For each of the test-particle atoms that reach our 
virtual detector, we now calculate $\omega$ as a result of a certain balance 
between the production and loss processes due to charge-exchange collisions 
between ISN He atoms and interstellar He$^+$ ions: 
\begin{equation}
	\frac{d \omega}{dt} = \betaprod(t) - \omega(t)\, \betaloss(t), 
\label{eq:defOmegaGainLoss} 
\end{equation} 
where the solution is carried out along a given atom trajectory and $\betaprod, \betaloss$ are the local instantaneous production and loss rates for the atoms following this trajectory. Since the atoms are following a Keplerian orbit, from conservation of angular momentum we can rewrite Equation~(\ref{eq:defOmegaGainLoss}) as a function of true anomaly $\theta$ of an atom on this trajectory. This requires a change of variables: $dt \rightarrow (r^2/L) d \theta$, where $L$ is the magnitude of angular momentum per unit mass of the given atom.

As initial conditions for Equation~(\ref{eq:defOmegaGainLoss}) we demand that 
\begin{equation} 
\omega(\theta_{\mathrm{src}}) = 
\frac{n_{\mathrm{He,LIC}}}{\left(\pi\, u_{\mathrm{T,LIC}}^3\right)^{1/2}}
\exp\!\left[-\left(\frac{\vec{v}_{\mathrm{src}} - \vec{u}_{\mathrm{LIC}}}{u_{\mathrm{T,LIC}}}\right)^2\right]. 
\label{eq:DefOmegaInitial} 
\end{equation} 
Here, $\vec{v}_{\mathrm{src}}$ is the 
velocity vector of the given atom at the locus where its trajectory intersects the 
heliocentric distance $r_{\mathrm{src}}$. At this point, the true anomaly of this 
atom is $\theta_{\mathrm{src}}$. With this initial condition, we solve the 
production and loss balance Equation~(\ref{eq:defOmegaGainLoss}) until the 
trajectory intersects the heliopause. The heliopause location is adopted directly from the heliosphere model. There, we register the statistical weight 
$\omega_{\mathrm{HP}}$ and we restart atom tracking along its trajectory towards 
the detector in order to obtain its survival probability against ionization losses inside the HP $w_{\mathrm{ion}}$, exactly as it was done originally by \citet{kubiak_etal:16a}. 
Effectively, an analog to Equation~(\ref{eq:omegaLocDef1}) becomes: 
\begin{equation} 
\omega_{\mathrm{LOC}}(\vec{r}_{\mathrm{obs}},\vec{v}_{\mathrm{obs}}) = 
\omega_{\mathrm{HP}}\, w_{\mathrm{ion}}; 
\label{eq:omegaLocDef2} 
\end{equation} 
note that both $\omega_{\mathrm{HP}}$ and $w_{\mathrm{ion}}$ depend on 
$\vec{r}_{\mathrm{obs}}$ and $\vec{v}_{\mathrm{obs}}$.

In principle, the simulated signal could be calculated using 
Equation~(\ref{eq:defSimSig}) taking into account all details of the measurement 
process and for all IBEX orbits for which observations of neutral He are 
available. In practice, however, solving for $\omega_{\mathrm{HP}}$ is a 
time-consuming numerical process and the problem becomes too demanding 
numerically. Therefore we simplify the simulation by calculating 
$F_{\mathrm{sim}}$ only for the middle time for several representative orbits; 
i.e., we do not perform averaging over good time intervals indicated in 
Equation~(\ref{eq:defSimSig}). Further simplifications are not needed due to the 
way the Warsaw Test Particle Model (WTPM) simulation package operates. We estimate the numerical precision of the model at $\sim 1$\%.

The gist of the novelty in our paper is employing the charge-exchange interactions 
in Equation~(\ref{eq:defOmegaGainLoss}) to calculate a balance between the gains 
and losses of the He atoms reaching the detector. The production and loss terms 
for the He atoms following a given trajectory are calculated assuming various 
models of the plasma flow between the calculation boundary and the heliopause. In 
all of the models considered, we assume that the interstellar matter includes the 
primary, unperturbed population of ISN He, which is not modified in any significant way in front of the heliopause and maintains its homogeneous velocity vector, 
temperature, and density. The interstellar matter includes a population of He$^+$ 
ions which are in equilibrium with the local plasma flowing past the heliosphere. 
The flow is assumed to be invariable with time. In the following section we 
present details of the production and loss terms used in the calculation.

The assumption that the density, bulk velocity, and temperature of neutral helium, used to calculate the production and loss terms in Equation~(\ref{eq:defOmegaGainLoss}), does not vary in the calculation region outside of the heliopause is quite realistic. Charge exchange with \hepl ions does not change the total number of neutral He atoms, and all He atoms in the region are potential candidates for charge exchange collisions, regardless of their collision history. Unlike in ISN H, modifications of the mean temperature and velocity of the total population of neutral He are very small because the absolute rate of charge exchange is small. The conclusion that the total He population very little differs from the unperturbed isothermal Maxwell-Boltzmann gas is supported by results of global heliospheric model with the helium charged and neutral populations taken into account self-consistently, shown in Figure~11 in \citet{kubiak_etal:14a}.

In our approach, we simulate the entire signal expected to be observed by IBEX 
without distinction what percentage of the simulated count rate in a given orbit 
and spin-angle bin is due to the pristine ISN He atoms that have not undergone any 
collision interactions, and which of the simulated counts are due to atoms 
produced by charge exchange collisions somewhere in front of the heliopause. 
Thus neither our simulation nor the observation is able to discriminate between atoms belonging to the primary and secondary populations.

This separation is not needed to compare the model with the observations. However, for illustration purposes, in some cases we will present the contribution of the secondary atoms to the simulated signal. This will be adopted as a difference between the simulated flux obtained for the given case (e.g., that presented in Section~\ref{sec:magnetizedFlow}) and that obtained for the hot model scenario (Section~\ref{sec:testCase}).

The question of definition of the secondary population is only seemingly simple. In fact, however, this question is quite complex and different authors adopt different definitions. \citet{osterbart_fahr:92}, for example, defined the secondary, tertiary, and higher-order populations as those composed of atoms that participated in one, two, or more collisions within the calculation region. \citet{mueller_etal:16a} defined it as that due to any charge-exchange collisions within the perturbed region beyond the heliopause and modelled it as a separate fluid. But what exactly is the perturbed region sometimes is not clear, especially with kinetic models. For example, \citet{zank_etal:13a} pointed out that when kinetic processes of interaction between the interstellar plasma and energetic neutral atoms emitted by the heliosphere are taken into account, then a region of increased plasma temperature extends far beyond the heliospheric bow wave or bow shock, and it is difficult to determine where exactly this heating effect vanishes. We alleviate all those dilemmas in our approach by adopting the working definition of the secondary component as specified in the preceding paragraph. If the calculation boundary is in a certainly undisturbed region, then calculating the target model and the reference hot model, and adopting the secondary population visible in the virtual collimator as the difference between these two is a well-defined and reasonable approach. The only differences between the solutions obtained from subtraction of the reference model from the target model can arise in the disturbed-plasma region, regardless how far from this region is the calculation boundary located. If in a given scenario there is no perturbation beyond the bow shock, still our secondary population is only due to the perturbed region. If, on the other hand, there is no bow shock or a perturbation beyond the bow shock exists, our definition does not change and the secondary population from the entire affected region is taken into account.

\begin{figure*} 
\centering 
\includegraphics[width=1.0\textwidth]{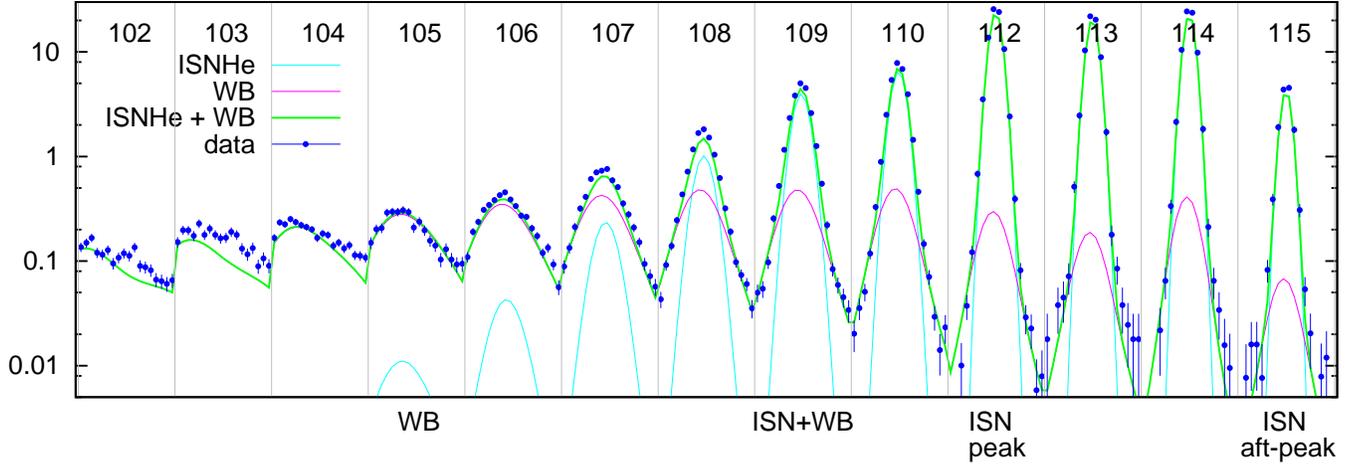} 
\caption{
Illustration of the orbits selected for this analysis against data set 
from the entire ISN observation season 2011. Data from individual orbits are presented 
between vertical bars, and the orbit numbers are shown below the upper frame. 
Orbit-averaged count rates are shown as blue dots with error bars. Each of the blue dots 
corresponds to data from an individual spin-angle bin. The magenta and cyan solid 
lines illustrate the model signal for the WB and ISN He populations, respectively, 
calculated for the parameters of ISN He obtained by \citet{bzowski_etal:15a} and 
those of WB obtained by \citet{kubiak_etal:16a}. In these calculations, all the 
details suggested by \citet{sokol_etal:15b} and \citet{swaczyna_etal:15a} were taken into account and it was assumed that the parent populations for ISN He and WB are separate collisionless, homogeneous Maxwell-Boltzmann distribution functions. The green line is the sum of the simulated WB and ISN He signals; note that for some orbits practically the 
entire signal is due to WB. The orbits selected for our discussion: 105 (WB), 109 
(ISN + WB), 112 (ISN peak), 115 (ISN aft-peak), are marked at the bottom of the 
upper panel.}
\label{fig:orbitSelection}
\end{figure*}

\subsection{Production and loss rates of He atoms in front of the heliopause} 
\label{sec:betaProdLoss}

As the production and loss processes we only admit the resonant charge-exchange 
collisions between He atoms and He$^+$ ions. However, it will be clear from 
further discussion that additional gain and loss processes may be added (e.g., 
elastic collisions, electron-impact ionization, recombination, photoionization, 
charge exchange with H atoms and protons). The approximations acceptable for use 
in the calculation of charge-exchange reactions for heliospheric hydrogen and 
proton populations were recently discussed by \citet{heerikhuisen_etal:15a}. They 
point out that for distribution functions in the Maxwell-Boltzmann approximation 
and moderate temperatures it is sufficient to approximate the rate of charge 
exchange $\beta_{\mathrm{AB}}$ of a particle of type A traveling at some speed 
relative to a stationary\footnote{i.e., with mean velocity equal to 0.} 
Maxwell-Boltzmann population of particles B with a density $n_{\mathrm{B}}$, 
temperature $T_{\mathrm{B}}$, and particle mass $m_{\mathrm{B}}$ by the formula: 
\begin{equation} \beta_{\mathrm{AB}} = n_{\mathrm{B}} \, u_{\mathrm{rel,AB}}\, 
\sigCX(u_{\mathrm{rel, AB}});\; u_{\mathrm{rel,AB}} = \urel(\vec{v}_{\mathrm{A}}, 
\vec{u}_{\mathrm{B}}, u_{\mathrm{T, B}}), \label{eq:simplifiedReactionRate} 
\end{equation} where $ u_{\mathrm{rel,AB}}$ is a certain effective relative speed 
between A and B. For this speed, we adopt: 
\begin{equation}
 u_{\mathrm{rel, AB}} = \left[\left(\VA +\frac{1}{2 \VA}\right) 
\mathrm{erf}(\VA)+\frac{\exp(-\VA^{\;2})}{\sqrt{\pi }} \right] u_{\mathrm{T, B}}, 
\label{eq:urelDef4} 
\end{equation} 
where $\VA = v_{\mathrm{A}}/u_{\mathrm{T,B}}$, 
and $u_{\mathrm{T,B}} = (2 k_B\, T_{\mathrm{B}}/m_{\mathrm{B}})^{1/2}$ is the 
thermal speed of population B, $\VA\, u_{\mathrm{T,B}}$ is the speed of particle A in the reference system co-moving with population B. The effective relative velocity $u_{\mathrm{rel, AB}}$ is in fact a function of four parameters: the mean velocity of population A relative to Sun, the mean velocity of population B relative to Sun, the thermal velocity of population B, and the specific velocity of a particle from population A with respect to its parent population. A formula for effective mean speed in this scenario was originally derived by \citet{fahr_mueller:67} and repeated by \citet{ripken_fahr:83a}. The formula we use, defined in Equation~(\ref{eq:urelDef4}), was used by \citet{fahr_bzowski:04b} and is very similar to that used by \citet{heerikhuisen_etal:15a}. With this formula, for 
relative speeds of A much larger than the thermal speed of B ($V_{\mathrm{A}} \gg 
1)$ the effective relative speed $\urel \rightarrow v_{\mathrm{A}}$, and for 
$V_{\mathrm{A}} \simeq 0$ the relative speed converges to the thermal speed: 
$\urel \rightarrow u_{\mathrm{T,B}}$.

We assumed that the charge exchange reaction defined in 
Equation~(\ref{eq:cxHe-Hepl}) operates without momentum change between the 
reaction partners. For the charge exchange rate we need a formula that does not show an 
unrealistic behavior for very low reaction speeds (close to 0) and is valid up to 
$\sim 100$~\kms. For this, we approximated a subset of data compiled by 
\citet{barnett_etal:90} by the following formula: 
\begin{equation} \sigCX 
\left(\urel \right) = 3.752\times 10^{-15}-5.212\times 10^{-16} \ln (\urel), 
\label{eq:sigmaDefinition1} 
\end{equation} 
where $\urel$ is the collision speed in \kms and $\sigCX$ is in cm$^2$.

The production rate $\betaprod (t)$ is determined by the probability (given by the 
value of the distribution function $f_{\mathrm{He}^+}$ of the interstellar \hepl 
ions) that a \hepl ion with the velocity vector $\vec{v}(t)$ exists in the 
population of \hepl ions at the location $\vec{r}(t)$ and that it enters into the 
c-x interaction with the ambient He population. This is reflected by the reaction 
cross section $\sigCX$ multiplied by the relative speed of the collision partners. 
This rate is approximated by the formula: 
\begin{eqnarray}
	\betaprod (t) &=&  \nHe\, \nHepl\, f_{\mathrm{He}^+}\!\left(\vec{v}, \vec{u}_{\mathrm{He}^+}, \uTHepl\right) \urel^{\mathrm{prod}}\, \sigCX \left(\urel^{\mathrm{prod}}\right);\nonumber \\
	\urel^{\mathrm{prod}} &=& \urel(\vec{v}, \vec{u}_{\mathrm{He}}, \uTHe) , 
\label{eq:defProdRate} 
\end{eqnarray} where we keep in mind that the ISN He 
distribution is assumed to be homogeneous in space and its density, velocity, and 
temperature do not vary with $\vec{r}$. The loss rate per He atom is proportional 
to the entire local density of \hepl ions because it is assumed that in the result 
of any c-x collisions the product He atom will have a different velocity vector 
from that before the collision. The loss rate per atom is then given by the formula: 
\begin{eqnarray} 
\betaloss (t) &=& \nHepl 
\,\urel^{\mathrm{loss}}\,\sigCX\left(\urel^{\mathrm{loss}}\right);\nonumber \\ 
\urel^{\mathrm{loss}} &=& \urel(\vec{v}, \vec{u}_{\mathrm{He}^+}, \uTHepl). 
\label{eq:defLossRate} 
\end{eqnarray} 
Note that the effective relative velocities 
$\urel^{\mathrm{prod}}, \urel^{\mathrm{loss}}$ in Equations (\ref{eq:defProdRate}) 
and (\ref{eq:defLossRate}) are different from each other.

\subsection{IBEX orbit selection}  
\label{sec:OrbitSelection}

Since on one hand our study has a qualitative character and on the other hand the 
amount of IBEX-Lo observations of neutral He is large, we believe it is not 
practical to simulate observations from all available orbits. Instead, we select 
four of them, characteristic for different aspects of the observed signal. We 
decided to choose data from the 2011 ISN observation season because during the 
first two seasons, 2009 and 2010, a considerable amount of ISN H was present in 
the data in addition to ISN He, which was reduced by a factor of 2 in the 2011 
season \citep{saul_etal:13a}. This was because due to an increase in solar 
activity in 2011, the solar radiation pressure increased and repelled most of ISN 
H atoms, preventing them from reaching the detector. Hence the counts observed in 
2011 by IBEX-Lo were almost entirely due to He atoms, both for the early orbits 
within the data set, where mostly WB is observed, and for the late, post-peak 
orbits, where \citet{saul_etal:12a} and \citet{saul_etal:13a} reported a 
considerable contribution from ISN H during the 2009 and 2010 ISN observation 
seasons.

The data from the 2011 ISN observation season are shown in 
Figure~\ref{fig:orbitSelection}. For our study we chose four representative orbits 
that are well suited to study various aspects of the global signal. We selected 
the orbit where the observed signal due to the pristine ISN He is maximum, i.e., 
orbit 112; an orbit where \citet{kubiak_etal:16a} suggested that only the WB 
signal is present, i.e., orbit 105; an orbit where the signal is due to both ISN 
He and WB in comparable proportions, i.e., orbit 109; and a late orbit during the 
season, where the signal is again almost entirely due to the ISN He population, 
and WB is expected to show up only far away from the signal peak, i.e., orbit 115. 
With this selection, we are able to study the symmetry of the simulated signal 
relative to the peak orbit depending on the adopted models of the ISN matter flow 
past the heliopause, to compare the intensity of the simulated signal with the 
data both for the peaks and in the far wings as a function of spin angle, and to 
study the buildup of WB in various scenarios.

\subsection{Models of the interstellar matter in front of the heliopause}
We performed simulations for several different scenarios for the interstellar matter 
in front of the heliopause, studying various aspects of the creation of the signal 
observed by IBEX-Lo. In all of the simulations, we assumed the densities of the neutral and charged H components in the unperturbed interstellar medium close to these obtained by \citet{bzowski_etal:08a} and \citet{bzowski_etal:09a} based on analysis of Ulysses pickup ion observations, and the temperature and bulk velocity of the unperturbed interstellar gas obtained by \citet{bzowski_etal:15a} and \citet{bzowski_etal:14a}, \citet{wood_etal:15a} based on direct-sampling observations on IBEX and GAS/Ulysses, respectively. The absolute density of ISN He far away from the Sun was adopted after \citet{witte:04} and \citet{gloeckler_etal:04b}, and the proportion between the neutral and two charged components of ISN He, used to calculate the absolute density of \hepl, after \citet{slavin_frisch:08a}. 

Note that in none of the adopted scenarios we assume preexistence of any 
population of neutral He in front of the heliosphere other than the unperturbed ISN He. 
All the differences in comparison with the classical hot model of ISN He are solely 
due to the action of charge exchange collisions between ISN He and interstellar 
\hepl ions.

\subsubsection{Hot model} \label{sec:hotModel}
We start from verifying the concept of the simulation and to 
that end we assume that the interstellar matter in front of the heliopause is not 
disturbed in any way by the presence of the heliosphere, i.e., that the plasma 
flow is homogeneous and isothermal. These assumptions are equivalent to the 
assumptions of the classical hot model of ISN gas distribution in the heliosphere, 
with the important exception that we allow for charge-exchange gains and losses in 
front of the heliopause. 

In this scenario, we expect to reproduce the signal due to ISN He, but not that 
of the WB.

\subsubsection{Bonn model} \label{sec:bonnModel}
\begin{figure} 
\includegraphics[width=0.5\columnwidth]{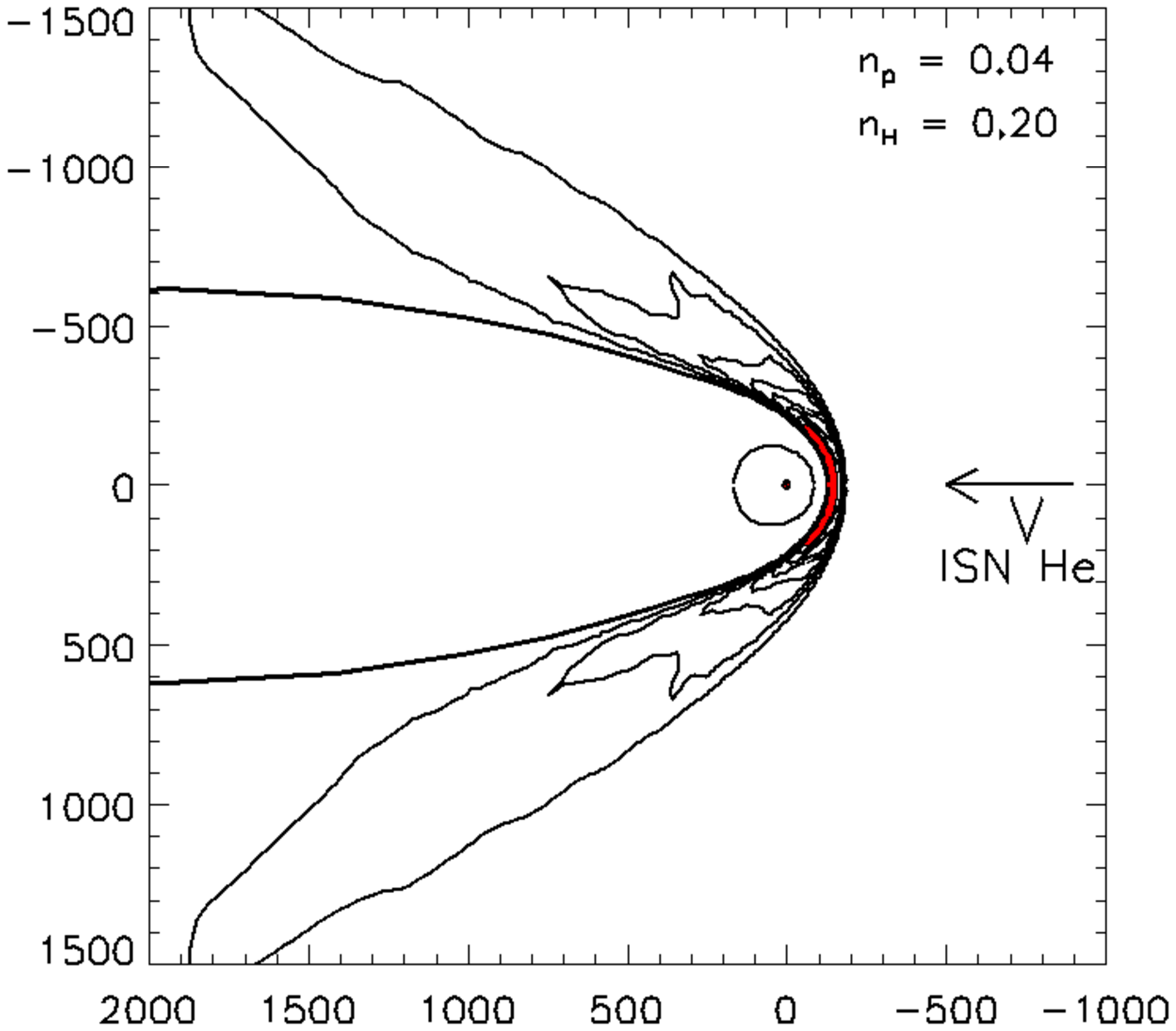} 

\includegraphics[width=0.5\columnwidth]{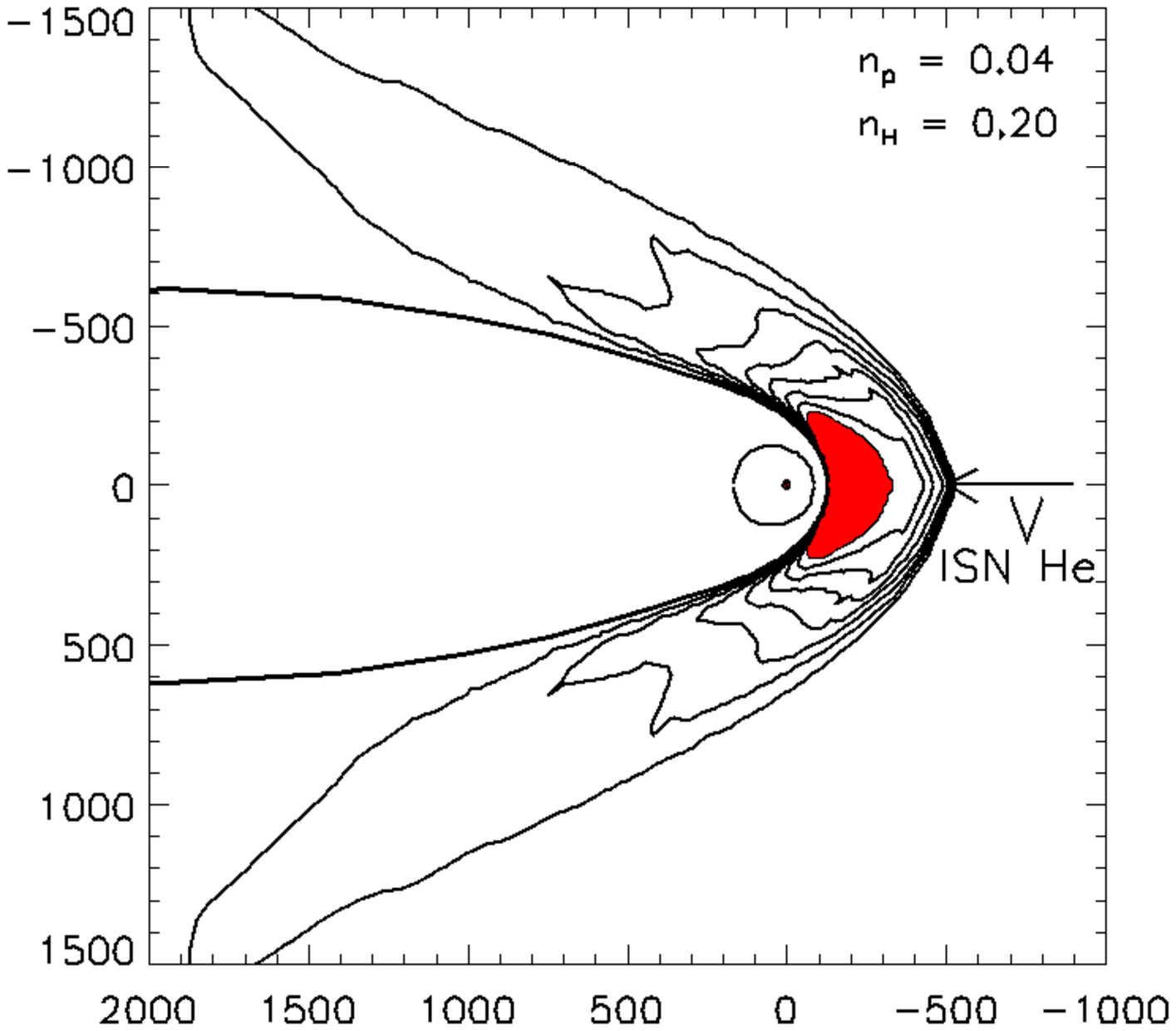} 
\caption{
Plasma density contours  
obtained from the axially-symmetric, hydrodynamic model of the outer 
heliosheath by \citet{kausch_etal:97a} (upper panel) and from the
modified version of the model, with the region between the bow shock
and the heliopause artificially expanded (lower panel).} 
\label{fig:model-bonn} 
\end{figure}

Subsequently, we do allow for modifications of the interstellar matter flow in the 
vicinity of the heliosphere but we neglect the interstellar magnetic field. 
In this case we can assume that the structure of the outer heliosheath is
axisymmetric with respect to the inflow direction of the interstellar medium.
For a detailed description we use the version of the Bonn model \citep{kausch_etal:97a},
based on a numerical solution of the hydrodynamical equations, with the 
inflow speed and temperature of the interstellar matter close to those obtained by 
\citet{bzowski_etal:15a}. 

To calculate the Breeze production rates we need the parameters of the \hepl component, 
which is not included explicitly in the Bonn model. We therefore assumed that the flow 
velocity and temperature of the \hepl ion component in the OHS are 
equal to the corresponding parameters of the (mainly hydrogen) plasma in the Bonn 
model, and that the \hepl number density is a given constant fraction of 0.15 of the bulk 
plasma density. Figure~\ref{fig:model-bonn} (top panel) shows the plasma density 
contours in the model.

We also considered a modified version of the model, in which spatial 
distributions of the parameters of the plasma and neutral components of the
outer heliosheath were artificially expanded in distance, in order to increase
the hot plasma region between the bow shock and the heliopause
(Figure~\ref{fig:model-bonn}, bottom panel).
The modified distributions $F'$ are related to $F$ of the original Bonn model by
$F'(r',\theta)=F(r,\theta)$ and $r'-r_{HP}(\theta)=\alpha(\theta)(r-r_{HP}(\theta))$, 
where $r$ is the heliocentric distance, $\theta$ the 
angle counted from symmetry axis, $r_{HP}(\theta)$ the distance to the heliopause,
and $\alpha(\theta)$ the $\theta$-dependent expansion factor. The modified model
is no longer a solution of the hydrodynamical equations, but can be used to 
provide an approximate estimation of the effect of a wide hot plasma region in
front of the heliopause. Effectively, this up-scaling results in a larger volume of the matter in front of the heliopause having an increased density and temperature.

In this scenario, we do expect some signal to appear in the IBEX 
orbit characteristic for WB (orbit 105) and we expect that the signal for the IBEX orbits where the primary ISN dominates to be modified due to the secondary population of He. In particular the heights of the peaks in orbits 109 and 115 will have different proportions to the height of the peak for orbit 112. Still, we expect systematic qualitative differences with the actual data because the flow past the heliopause in this scenario has axial symmetry around the inflow direction, which is not expected in reality due to the action of ISMF. By comparing two very similar models, differing only by the size of OHS, we can study the effect of the size of the OHS (equivalently, the ``optical depth'' for gains and losses) on the WB signal. An important aspect of this modified model is that the region of a higher density and temperature of the plasma is significantly larger than previously.

\subsubsection{MHD model} \label{sec:MHDmodel}

\begin{figure} 
\centering 
\includegraphics[width=1.0\columnwidth]{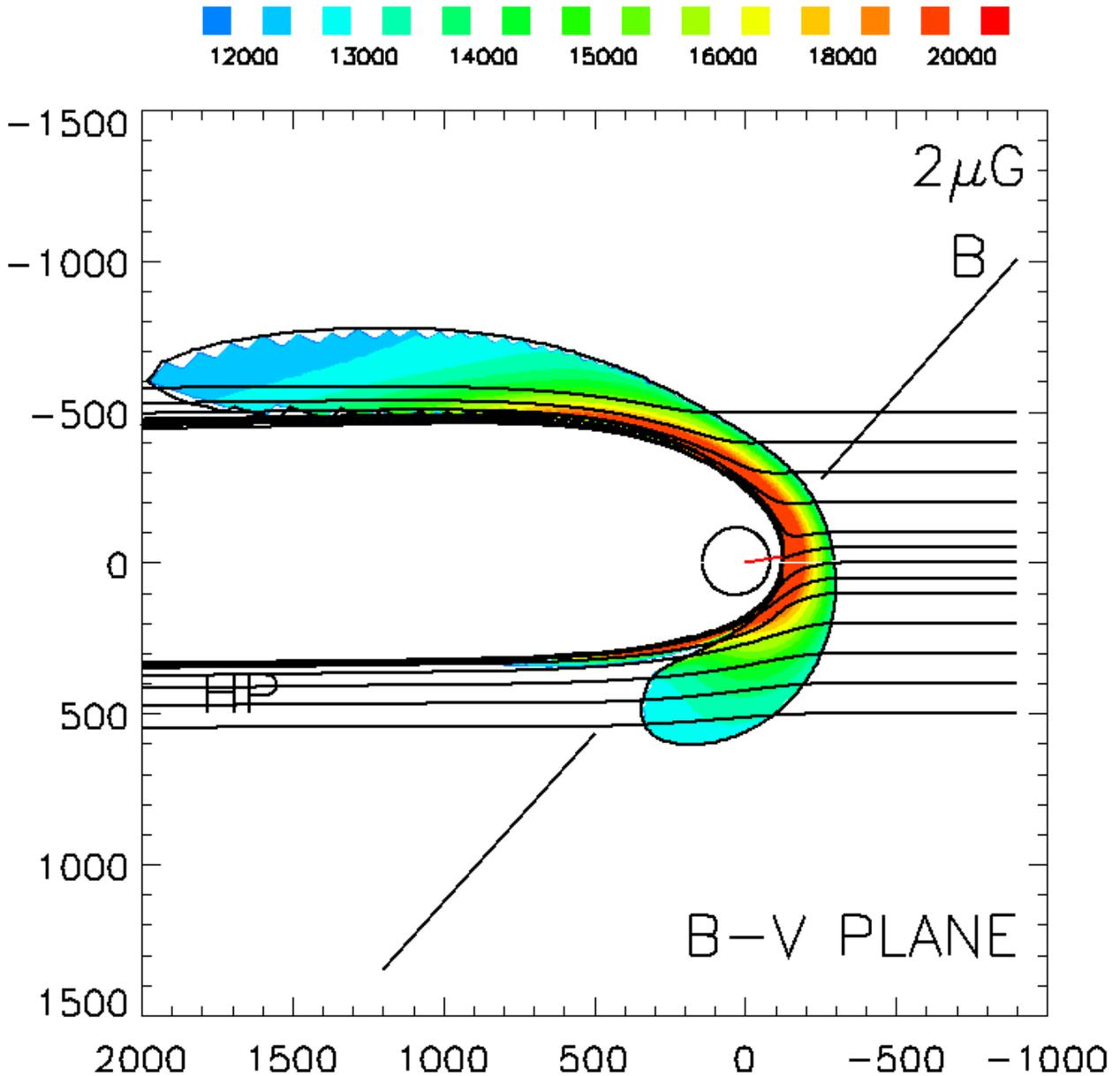} 
\caption{
Color-coded distribution of plasma temperature within the region 
of enhanced plasma density outside the heliopause in the 
B-V plane (the plane containing the Sun, the interstellar magnetic field 
and the interstellar matter velocity direction). The temperature scale 
corresponds to the modified (rescaled) model. 
The unperturbed interstellar 
matter flows in from the right-hand side and its flow lines within the outer 
heliosheath are drawn with the black solid lines. The direction of the magnetic 
field is shown by the line denoted B. This direction was adopted towards the 
center of IBEX Ribbon. The small black oval centered at (0,0) is the termination 
shock. The inflow direction of WB found by \citet{kubiak_etal:16a} is shown by the 
short red line extending from the Sun. } 
\label{fig:2muG-temp-streamlines} 
\end{figure}

Finally, we take into account the interstellar magnetic field, which breaks the 
axial symmetry of the interstellar flow past the heliopause, and we consider the 
role of plasma temperature. We use the results of the MHD simulations 
described in \citet{grygorczuk_etal:14a} and \citet{czechowski_etal:15a}, for the particular 
case of interstellar field strength of 2~$\mu$G. The velocity vector and 
temperature of ISN gas far away in front of the heliosphere are taken from 
\citet{mccomas_etal:15a} and the direction of the interstellar magnetic field 
towards the IBEX Ribbon center from \citet{funsten_etal:09b}. The MHD code used in 
our calculations is a corrected version of the one described by 
\citet{ratkiewicz_etal:08a}. 

Our MHD model includes the interaction of plasma with the neutral 
component of the interstellar medium in a simplified form, in which 
the neutral flow is assumed to be constant, and there is no feedback 
from the plasma to ISN H.  As a result of this simplification, the plasma temperature obtained 
tends to be lower than in global heliospheric models which take this 
feedback into account. Therefore, we consider two versions of the model: 
the MHD solution, and a modified version that differs from the MHD
solution by the artificially increased temperature (multiplied by 2) in the 
region inside the isodensity contour 0.067~cm$^{-3}$), which was chosen to define the boundary of the high plasma density region immediately outside the heliopause. The question of the temperature of the interstellar \hepl component of the OHS plasma is not trivial anyway. As pointed out by \citet{mueller_etal:16a} in the context of creation of the secondary helium component, available literature emphasizes  that charge exchange heats plasmas \citep{zank_etal:96a, zank_etal:13a} and that the long equilibration time scales may result in the lack of equilibrium between the ions of different species in OHS \citep{zank_etal:14b}.

Figure~\ref{fig:2muG-temp-streamlines} shows the heliopause, termination shock,
the color-coded temperature distribution inside the 0.067~cm$^{-3}$ plasma isodensity 
contour, and selected plasma flow lines in the B-V plane (the plane containing 
the Sun, the interstellar magnetic field and the interstellar matter velocity 
direction), which is the symmetry plane of the MHD solution. The B-V plane contains the 
inflow direction of WB found  by \citet{kubiak_etal:16a}), shown here by 
the short red line extending from the Sun. The temperature distribution 
color bar scale corresponds to the modified (doubled) values.
The temperature for the nominal model can be read from the same figure
with the scale values divided by two.

The helium component is not included explicitly in the MHD model. We obtain
the density, velocity and temperature of the \hepl ions based on the appropriate values of the simulated H$^+$ plasma using the procedure described above for the Bonn model.  

In this scenario, we may hope to see the features of the simulated signal close to the data because the OHS model has, at least approximately, the essential features expected in the reality, i.e., the decoupling of the plasma flow from the neutral gas flow, compression and heating of the plasma, and an asymmetry of the flow due to the action of ISMF. With two versions of the temperature, we can study the effect of the temperature on the WB signal. 

\section{Results}  
\label{sec:results}

\subsection{Test case: the conditions characteristic for the classical hot model 
of ISN He in the heliosphere} \label{sec:testCase}

\begin{figure*} 
\centering 
\includegraphics[width=1.0\textwidth]{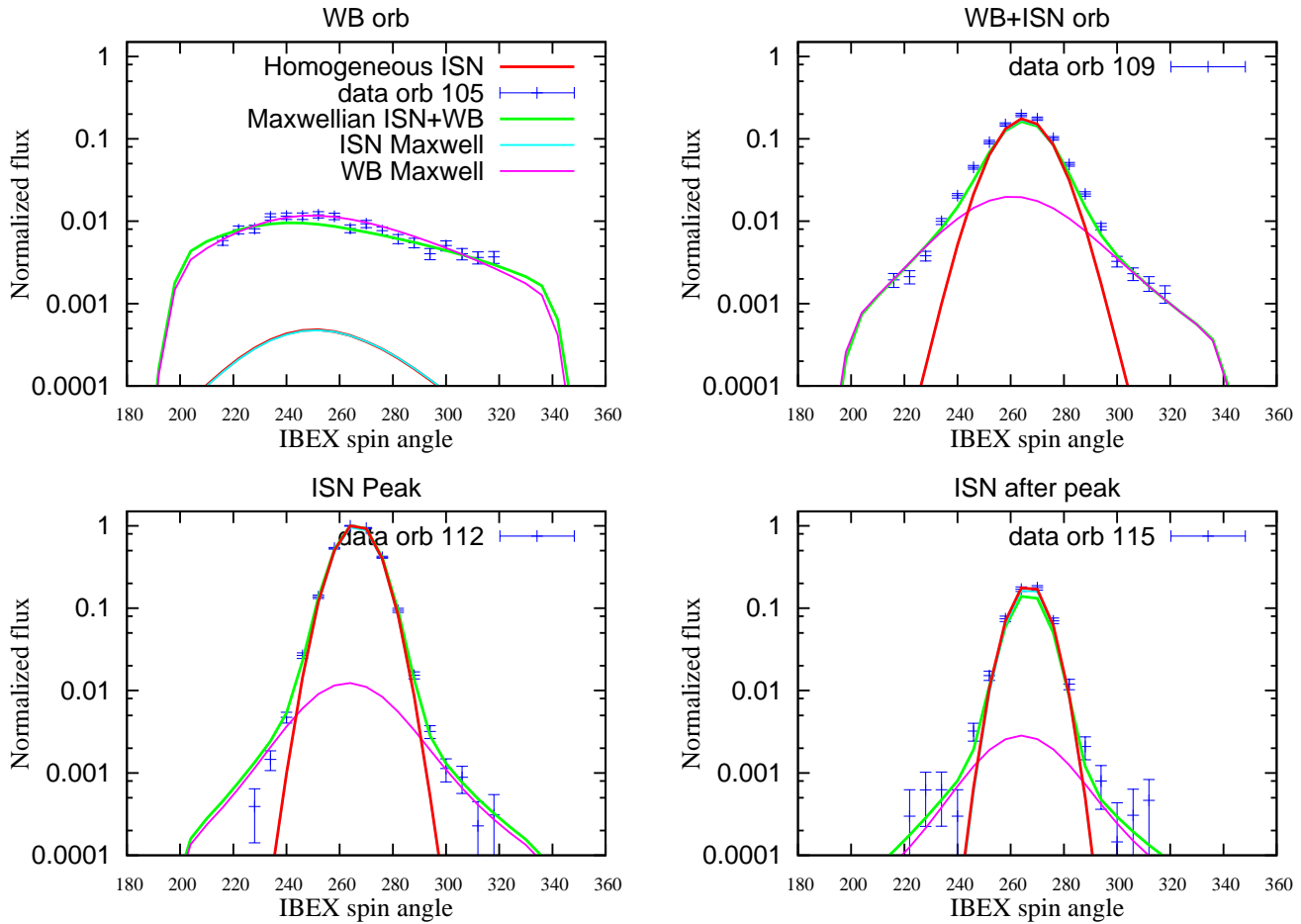} 
\caption{
Comparison of (1) the data (blue dots), (2) model used 
by \citet{kubiak_etal:16a}, composed of the Maxwell-Boltzmann populations 
corresponding to ISN He (cyan) and WB (magenta), (3) the sum of the latter two 
(green), and (4) results of the present simulation for the homogeneous interstellar 
gas (red). The horizontal axis corresponds to the spin angle. Both the data and the models  were normalized to their respective maximum values for orbit 112.} 
\label{fig:hotModel} 
\end{figure*}

Results of the simulation carried out assuming that both plasma and ISN gas are 
unperturbed everywhere beyond the heliopause, but that charge-exchange collisions operate outside the heliopause and contribute to the gains and losses of the He atoms following trajectories that hit the 
detector, are shown in Figure~\ref{fig:hotModel} (red line). The signal characteristic for the 
unperturbed ISN He population is reproduced well, and the signal attributed by 
\citet{kubiak_etal:16a} as due to the Warm Breeze does not show up, in agreement 
with expectations. For the WB orbit (105) there is a very small signal predicted 
by both \citet{kubiak_etal:16a} and the present model. Both of them, however, are 
below the background in the data, which for this orbit is at a level of $\sim 10^{-3}$ 
\citep{galli_etal:15a}. For the orbits where ISN He dominates, WB is well 
visible at the wings of the central cores in the data and in the model by 
\citet{kubiak_etal:16a} but not in the present simulations.

These results suggests that (1) the baseline concept of the simulation including 
the balance between the charge exchange gains and losses in the outer heliosphere 
is correct and our numerical code does not produce artefacts, and (2) indeed, charge exchange collisions in front of the heliosphere 
are not able to produce a signal due to He atoms in the region attributed to WB, 
when the plasma and gas in front of the heliosphere are unperturbed and 
thermalized with each other. In reality, the existence of these conditions is 
not expected.

\subsection{Axially-symmetric flow in the absence of interstellar magnetic field
(Bonn model)} 
\label{sec:axiallySymmetricCase}

\begin{figure*}
\centering
\includegraphics[width=\textwidth]{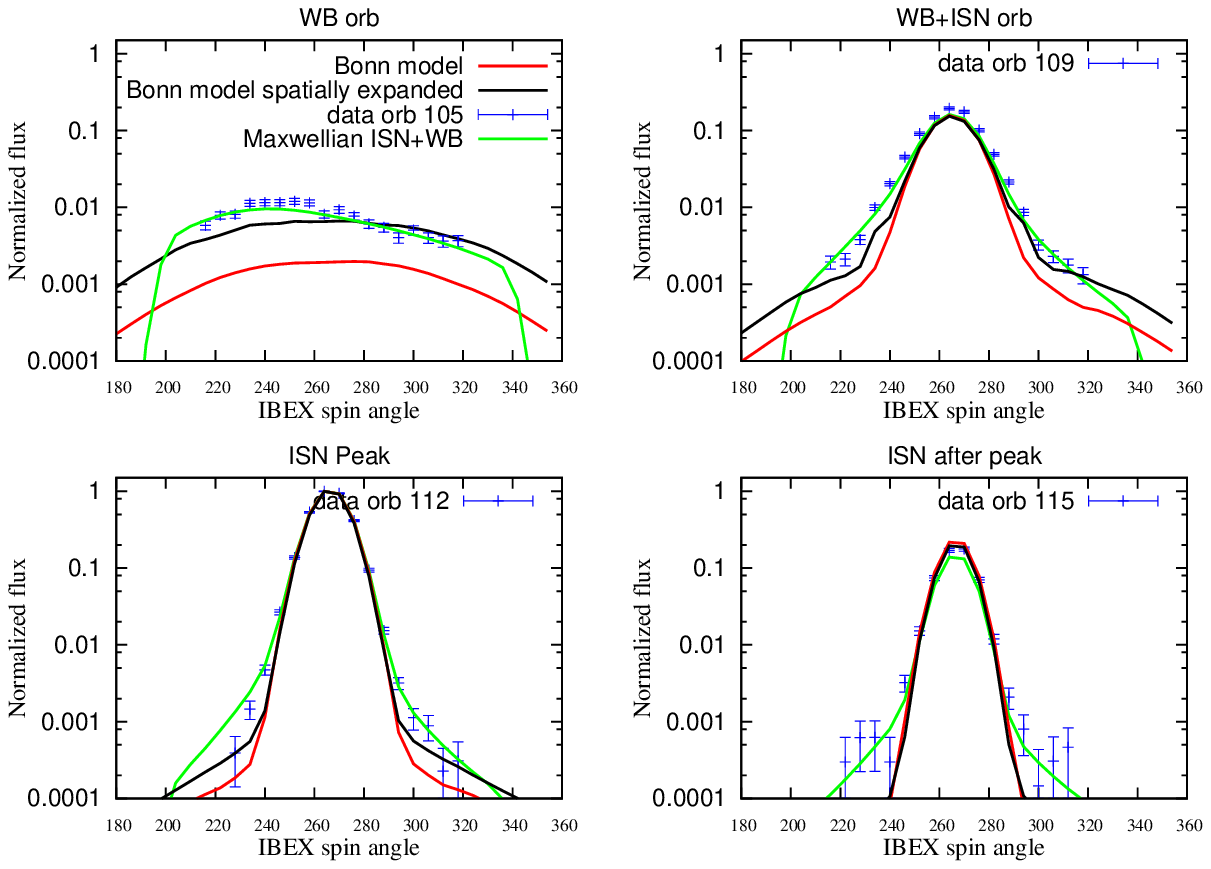}
\caption{
Comparison of (1) the data (blue dots), (2) model ISN + WB used by \citet{kubiak_etal:16a} 
(green), (3) results of the present simulation with the plasma flow around the 
heliosphere according to the Bonn  model (red; cf upper panel of 
Figure~\ref{fig:model-bonn}) and (4) results of the present simulation for the 
plasma flow from the re-scaled Bonn model (black; cf the lower panel of 
Figure~\ref{fig:model-bonn}). The horizontal axis corresponds to the spin angle. 
Both the data and the models were normalized to their respective maximum values for 
orbit 112. }
\label{fig:fluxes-log2-Bonn}
\end{figure*}

Results of this simulation suggest that when the plasma flow in front of the 
heliopause is disturbed, a signal in the region characteristic for WB does appear. 
It is visible both in the pure-WB orbit 105 and in 
orbits 109 and 112 (as elevated wings), but not in the after-peak orbit 115, as illustrated in Figure~\ref{fig:fluxes-log2-Bonn}, red line. However, the reproduction 
of the actually observed signal is not satisfactory even qualitatively. The 
maximum of the simulated signal for orbit 115 is too high by almost 30\% 
and for orbit 109 too low by $\sim 25$~\%. For the 
pure-WB orbit 105, the signal is too weak by an order of magnitude.

The quality of approximation given by the axially symmetric model is improved when 
the spatial extent of the OHS is artificially expanded, so 
that OHS extends to $\sim 450$~AU from the Sun, shown in the lower panel in 
Figure~\ref{fig:model-bonn}. The simulated signal is presented by black line in 
Figure~\ref{fig:fluxes-log2-Bonn}. The wings of the signal are now elevated and 
they are qualitatively similar to the observed ones for orbits 109 and 112, and 
the signal is at a level well comparable to the observed one for the pure-WB orbit 
105. However, the asymmetry in the peak heights between the orbits remains.

These results suggest that the intensity of the Warm Breeze signal is a function 
of the linear size of the outer heliosheath, as may be intuitively expected. Which of the parameters of the plasma is decisive in the reproduction of the observed signal: plasma density or temperature, will become evident in the next section. The systematic difference between the maxima of the model signal and that actually measured in the pre- and after-peak orbits suggests that axially symmetric heliospheric models are not likely to match the observed signal because the heliosphere is expected to be distorted due to interstellar magnetic field.

\subsection{Asymmetric flow by the magnetized plasma (MHD model)} 
\label{sec:magnetizedFlow}

\begin{figure*} 
\centering 
\includegraphics[width=1.0\textwidth]{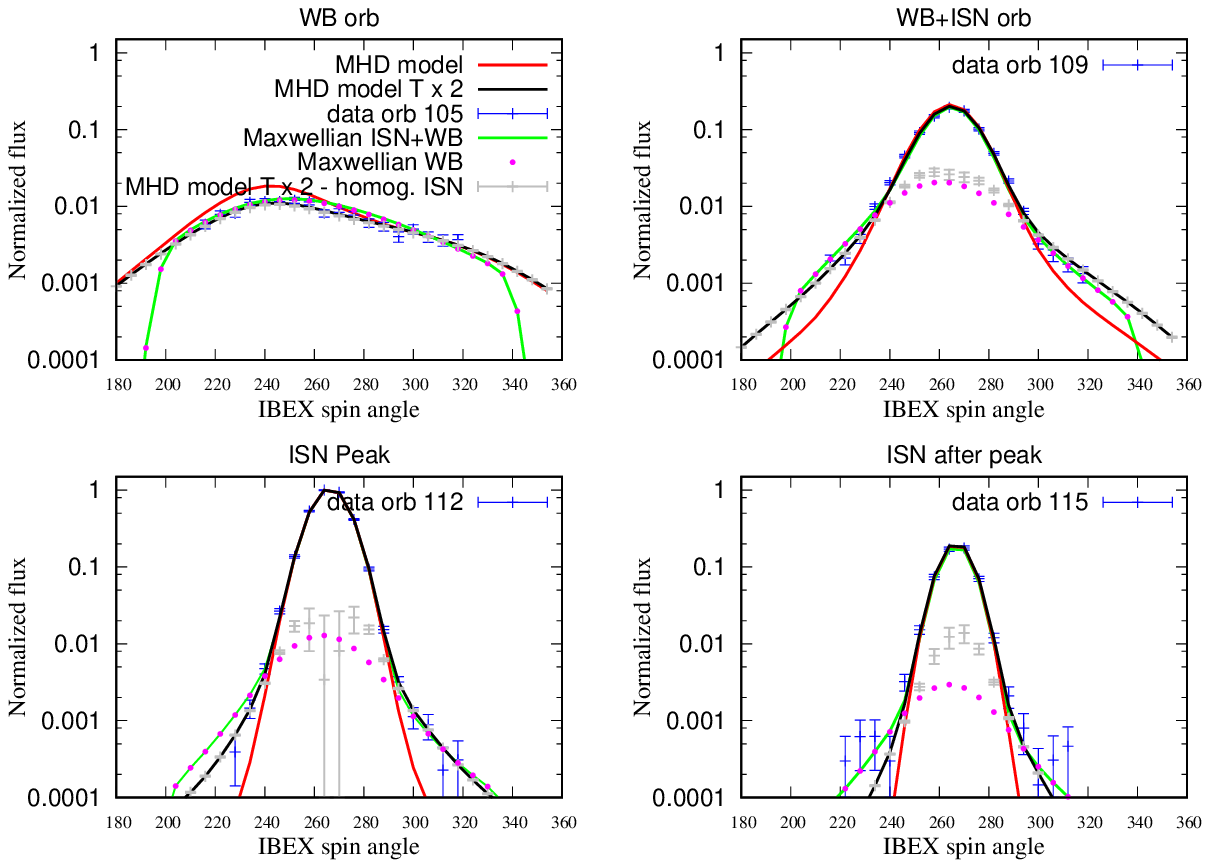} 
\caption{
Comparison of the data (blue dots) with the ISN + WB model from \citet{kubiak_etal:16a} (green) and simulation results for the MHD model of the outer heliosheath for interstellar magnetic 
field direction towards the center of IBEX Ribbon and an intensity of 2~$\mu$G. 
Red lines mark the simulations performed for the nominal MHD model, black lines for the modified version of the model (Figure~\ref{fig:2muG-temp-streamlines}). Purple dots mark the Warm Breeze given by the Maxwell-Boltzmann population with the parameters found by \citet{kubiak_etal:16a} and gray dots with error bars illustrate the difference between the modified MHD model (black line) and the hot model with no plasma modifications in front of the heliopause (the model drawn with red line in Figure~\ref{fig:hotModel}). The error bars mark the numerical precision of the difference of the two models. The horizontal axis corresponds to the spin angle. Both the data and the models were normalized to their respective maximum values for 
orbit 112.} 
\label{fig:fluxes-log2-002sym} 
\end{figure*}

Results of simulating the neutral He signal observed by IBEX in the scenario with the ISMF included in the MHD model are presented in Figure~\ref{fig:fluxes-log2-002sym}
(red lines for the nominal case, black lines for the modified model with an increased temperature). 

For the nominal MHD scenario, the signal for the WB orbit 
105 fits qualitatively quite well to that observed. The heights of the signal 
peaks for the pre-peak and after-peak orbits 109 and 115 also agree very well with 
those observed (note that in this simulation we did not average the simulated flux 
over the good times, as it had been done by \citet{kubiak_etal:16a}). This 
suggests that the distortion of the heliosphere by the interstellar magnetic field 
with the direction and strength suggested by recent analysis 
carried out using state of the art models \citep{zirnstein_etal:16b} is needed to 
account for the signal from neutral He observed by IBEX. On the other hand, not 
all aspects of the actually observed signal were reproduced: the wings of the 
signal for all orbits except 105 were obtained too low.

For the modified MHD scenario, increasing the plasma temperature resulted in an 
increase of the simulated signal in the wings without affecting the peaks, so that now also this aspect is in a qualitative agreement with observations. This illustrates the sensitivity of the system to various aspects of the outer heliosheath, including its asymmetry of the 
flow due to magnetic field, spatial extent of the outer heliosheath, and particularly the 
temperature of plasma. 

Ultimately, a simulation of the plasma flow in the OHS obtained from a relatively simple MHD model (with some temperature modifications) reproduced the signal due to neutral He observed by IBEX with a surprisingly good accuracy, at least for the selected representative subset of data. These simulations were carried out assuming the most up-to-date results for the direction and strength of LIMF and of the flow direction, velocity and temperature of interstellar gas in the LIC, and with the charge-exchange coupling between the \hepl and He components of interstellar matter. 

Keeping in mind the difficulty in formulating a clear definition of the secondary population in our approach, as discussed at the end of Section~\ref{sec:SignalSimulation}, we define the secondary population seen by IBEX as the difference between the flux simulated for a given test case and the flux simulated for the case with no plasma disturbance in front of the heliopause, discussed in Section~\ref{sec:hotModel} and shown in Figure~\ref{fig:hotModel}. The secondary population signal thus defined is presented with the gray dots in Figure~\ref{fig:fluxes-log2-002sym}. For comparison, we additionally show the secondary population approximated by \citet{kubiak_etal:16a} as due to a homogeneous Maxwell-Boltzmann flow outside the heliopause. Not surprisingly, in the pure-Warm Breeze orbit 112 the secondary population is responsible for practically entire signal, and for the pre-peak and peak orbits 109 and 112, respectively, it is close to the predictions given by the Maxwell-Boltzmann case. Note, however, that the numerical accuracy of our simulations, which is $\sim 1$\%, does not allow to separate the secondary population very accurately for the peak orbit 112. For the after-peak orbit 115, the model suggests the secondary population is much more plentiful close to the peak of the signal than in the two-Maxwellian case from \citet{kubiak_etal:16a}. We reiterate, however, that it is the agreement of the entire simulated signal with the data that is decisive to adopt or reject a given model. 

\subsection{In search for filtration of the primary ISN He population observed at 
1~AU} \label{sec:filtration}

Finally, we have assessed the hypothetical effect of filtration of the unperturbed 
ISN He population in the outer heliosheath. This effect was supposed by several 
authors \citep[e.g.][]{mobius_etal:15b} to bias the ISN He parameters determined 
from direct-sampling observations of ISN He atoms, in particular the temperature 
of ISN He gas in front of the heliosphere. It was hypothesized that the beam of ISN 
He observed at $\sim 1$~AU from the Sun may be systematically narrowed to some 
extent as a function of spin angle, because some of the original interstellar 
atoms could be preferentially ionized in the outer heliosheath and thus eliminated 
from the observed sample. This effect could show up as a reduction of the width of 
the observed beam. Neglecting this hypothetical effect could lead to a biased 
estimate of the temperature of ISN He gas ahead of the heliosphere. This effect could 
be expected in analogy to the behavior of the primary and secondary populations of ISN H, 
suggested by \citet{izmodenov_etal:03b} and \citet{izmodenov_etal:03a} based on the Moscow MC model of the heliosphere. 

With the simulations presented in this paper we are able to investigate at least 
qualitatively this hypothetical filtration effect. For comparison we took the 
results of the model where no plasma modification in OHS is considered on one hand 
(Figure~\ref{fig:hotModel}), and on the other hand the model that fits best to the 
observations, i.e., the model with the ISMF intensity 2~$\mu$G and the increased plasma temperature 
(Figure~\ref{fig:fluxes-log2-002sym}, black line). The first of these two models 
corresponds to the assumptions of the unperturbed hot model of ISN He in the 
heliosphere, similar to that used for analysis of the primary ISN He population by 
several authors \citep[e.g., ][]{witte:04, bzowski_etal:12a, mobius_etal:12a, 
bzowski_etal:14a, bzowski_etal:15a, leonard_etal:15a, mobius_etal:15b, 
schwadron_etal:15a}. The other one represents a more realistic physical scenario, 
where the gains and losses due to charge exchange in OHS are taken into account. 
We compared the simulated flux for the peak and after-peak orbits, for which the 
WB is expected to be visible only in the far wings of the observed signal, and we 
took for analysis the spin angle range exactly as that used in the analysis of ISN 
He by \citet{bzowski_etal:15a}, i.e., $\psi \in (252\degr, 288\degr)$. The 
simulated signals were fitted with a Gaussian function plus a constant background: 
\begin{equation} 
f(\psi) = bkg + f_0 \exp\left[-\left(\frac{\psi - \psi_0}{\sigma} 
\right)^2 \right] 
\label{eq:GaussModel} 
\end{equation} 
and the values of $\sigma$ 
obtained were $\sim 10.0\degr$ for the peak orbit and $\sim 8.7\degr$ for the 
after-peak orbit for the case of the classical hot model.

We found that consistently for both orbits considered, the $\sigma$ parameter fit 
to the classical hot model was a little {\em lower} than that obtained for the model 
where the OHS c-x collisions were allowed for. These differences were very small 
in comparison with the $\sigma$ value, $0.25\degr$ and $0.16\degr$ for the peak 
and after-peak orbits, respectively, i.e., about 2\%. Therefore any expected 
systematic bias due to the neglected charge-exchange collisions operating in the 
OHS, affecting the temperatures found by the above-mentioned authors, is of a 
similar magnitude, i.e., negligible in comparison with the reported error margins. 
An interesting detail, however, is that our simulations suggest that the small 
systematic effect is opposite to that hypothesized: instead of cooling the ISN He 
beam by differential ionization, some heating is observed, i.e., there is a net 
production of neutral He atoms observed in this interval of IBEX spin angles. This result indicates that the secondary component contributes to the entire signal observed by IBEX, including the pixels that in the original analysis of ISN He were treated as containing the pure ISN He gas. For the several pixels near the maximum count rate, this contribution is, however, very small and practically does not affect the width of the observed neutral helium beam.

\subsection{Source regions for the He atoms observed by IBEX in different regions in 
the sky} \label{sec:srcRegionSky}

\begin{figure*}
\centering
\includegraphics[width=0.4 \textwidth]{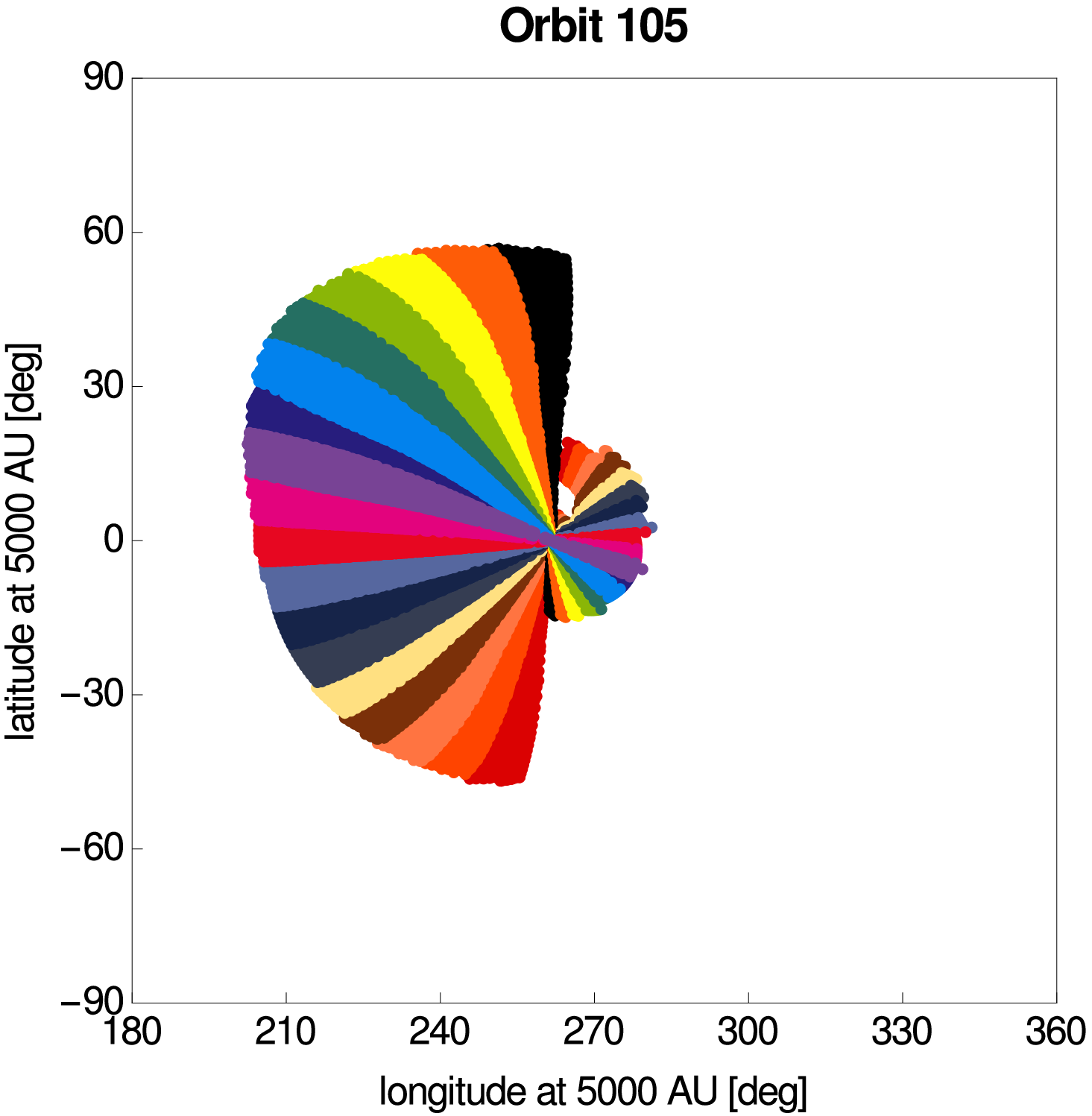}
\includegraphics[width=0.4 \textwidth]{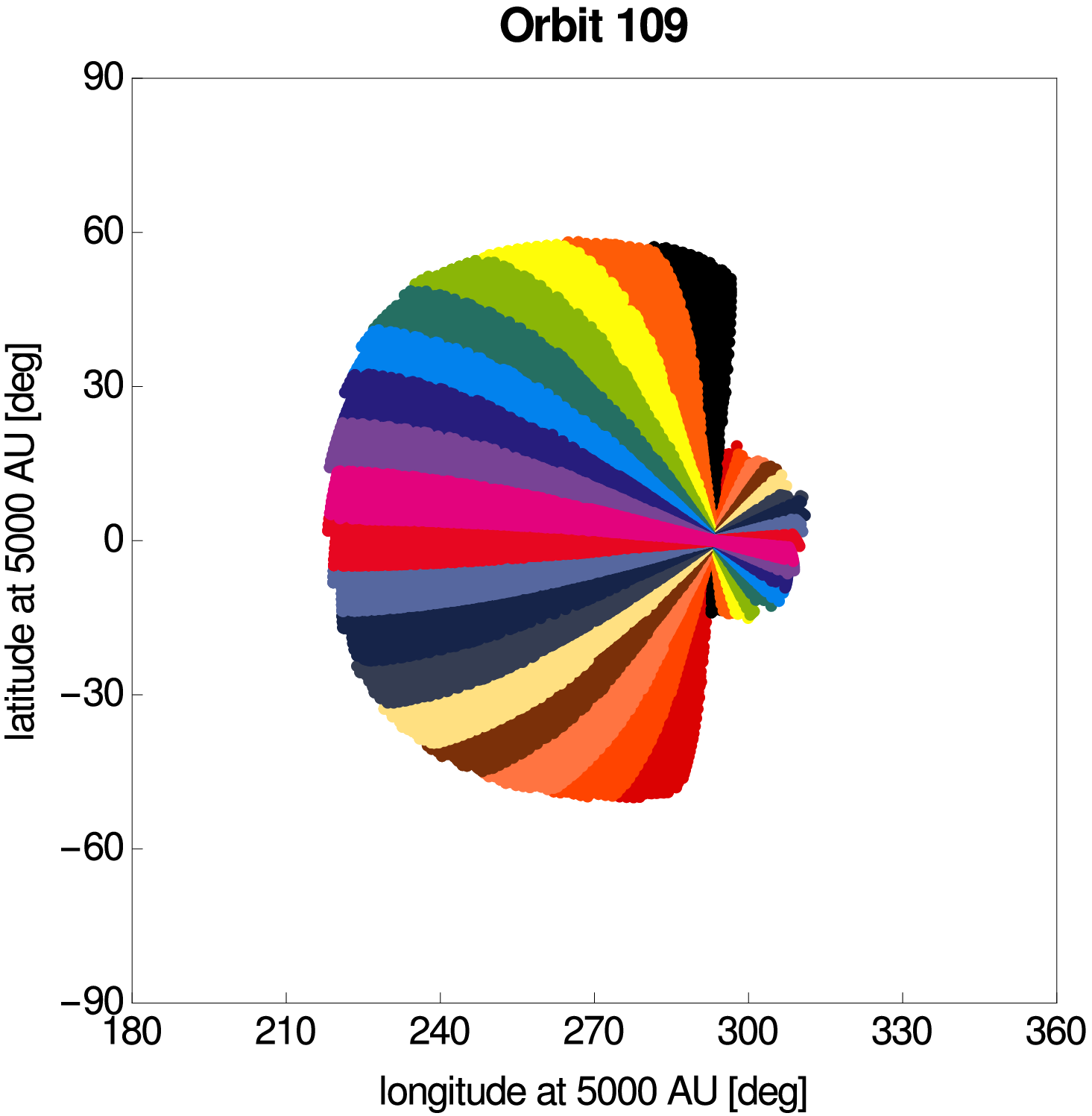}

\includegraphics[width=0.4 \textwidth]{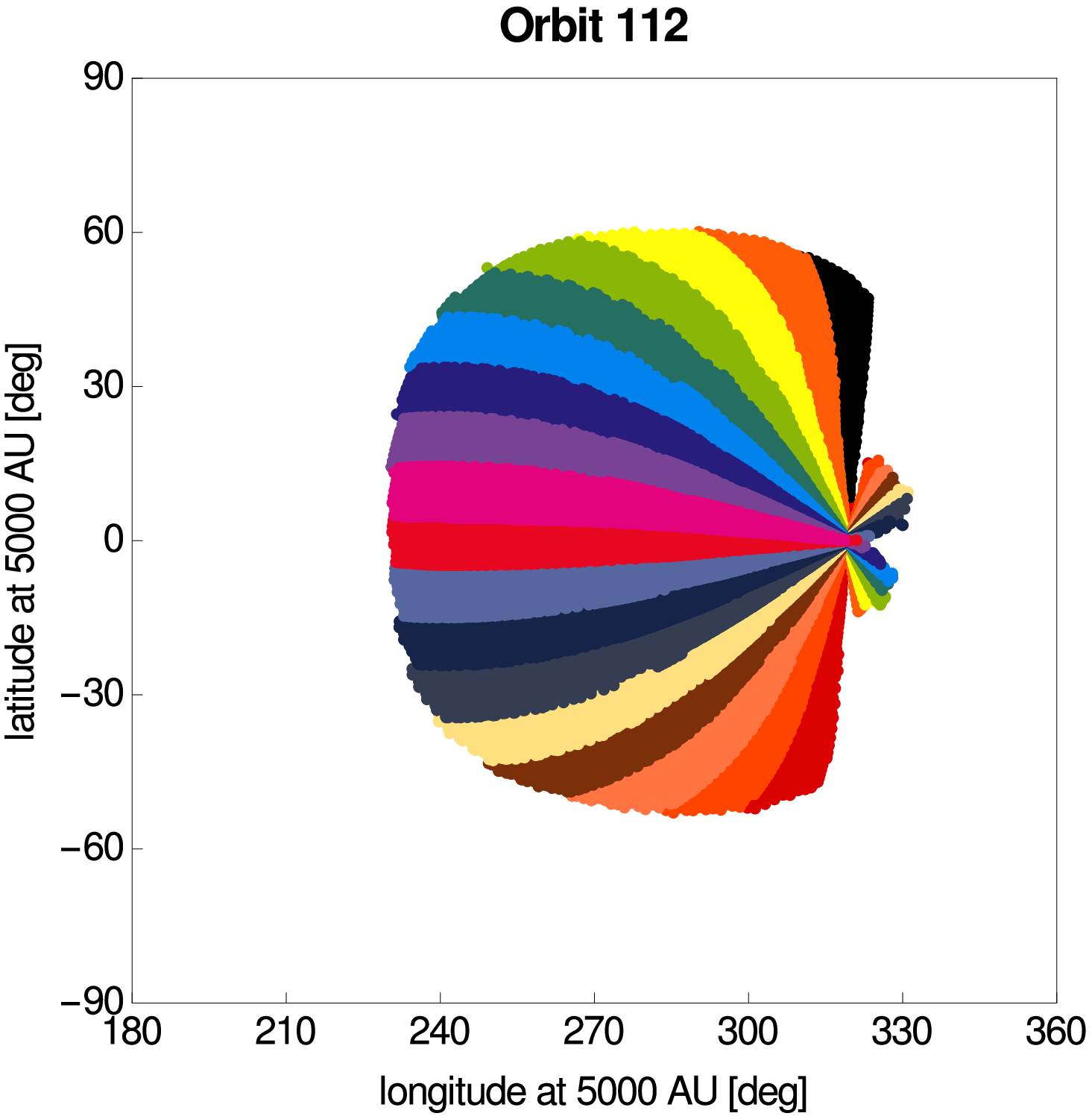}
\includegraphics[width=0.4 \textwidth]{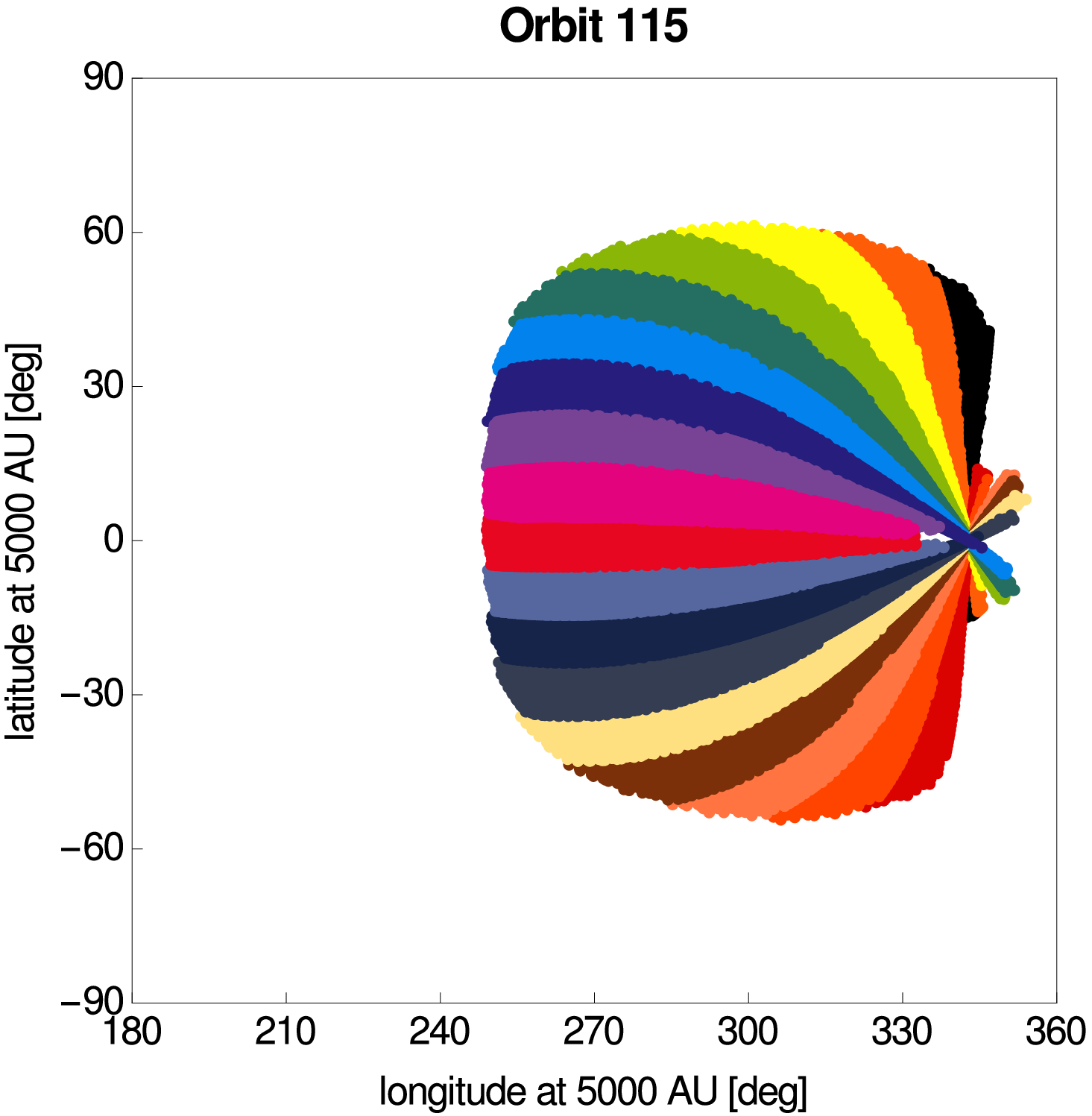}

\includegraphics[width=0.9 \textwidth]{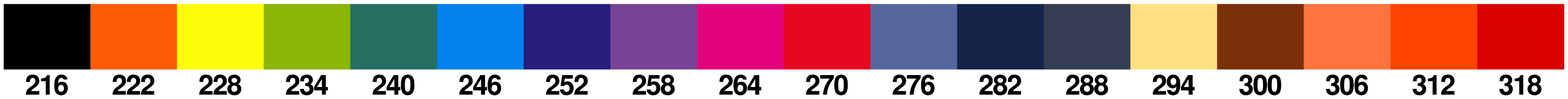}
\caption{
Ecliptic coordinates of the regions at boundary of the simulation region at 5000~AU from the Sun, where the atoms forming the signal observed by IBEX-Lo in individual pixels enter the calculation. The color code for spin angle bins (pixels) is given in the color bar. The spin angle bins correspond to the bins used to plot the simulated signal in Figure~\ref{fig:fluxes-log2-002sym}. The atoms used to determine these regions are those that contribute at least 0.001 of the calculated total flux in a given pixel.}
\label{fig:srcRegions}
\end{figure*}

A careful analysis of the ballistics of He atoms contributing to the total signal observed by IBEX at individual orbits showed that atoms registered in individual pixels enter the simulation region and cross the heliopause in well-defined strip-like regions, as illustrated in Figure~\ref{fig:srcRegions}. The atoms that define those regions are selected as those that contribute at least 0.001 of the total flux in a given pixels, but we verified that the regions practically do not change when one increases the cut-off level to 0.01 or 0.1. The overlap between the regions corresponding to individual pixels is approximately 20\% of the widths of the strips. This overlap does not change with a change in the cut-off level. The exact location of the strips depends on one hand on the time of observation (the ecliptic longitude of the spacecraft), and on the other hand on the exact pointing of the spacecraft spin axis. However, the yearly pattern is approximately repeatable. The heliosphere entry regions for atoms contributing to the signal in a given pixel (spin angle bin) moves from orbit to orbit, but for a given orbit and a given pixel the entry region is definitely identifiable.

A similar observation was made by \citet{kubiak_etal:14a} in the WB discovery paper, where the two Maxwell-Boltzmann populations model was used. The existence of the well-defined entry regions at the heliopause for atoms observed in individual spin angle bins is due to ballistics. Hence these regions are relatively robust against various assumptions on the processes operating in the OHS. Since most of the atoms follow approximately straight-line trajectories prior to entering the heliosphere, the regions identified in Figure~\ref{fig:srcRegions} extend radially away without significant flexing at least for a few hundred AU from the heliopause. Thus the signal observed in individual pixels is an integrand over well-defined regions in the three-dimensional space outside the heliopause and this fact can be used to test various hypotheses on the behavior of matter in the OHS. 

\subsection{Outlook} \label{sec:outlook}
The good qualitative agreement between the simulation results and the subset of observations we had chosen for comparison suggests that the methodology of simulating the expected signal has large diagnostic potential and may be used to discriminate between qualitatively different models of the heliosphere. Nowadays, three different models of the heliosphere are discussed in the literature: the canonical comet-like model, similar to that used in this paper \citep[e.g., ][]{izmodenov_alexashov:15a, heerikhuisen_etal:14a, pogorelov_etal:09b}, a croissant-like model due to \citet{opher_etal:15a}, or the diamagnetic bubble-like hypothesis due to \citet{krimigis_etal:09a}. A simulation of the expected signal of the secondary He population resulting from these models, calculated for any physical assumptions and values of the relevant parameters, can be done relatively easily based on a prediction of the plasma streamlines, density and temperature in a region in front of the heliopause. 

Another logical step forward would be testing predictions of the current model against the entire data set available. This, however, requires optimizing the code to have it run faster. This work is underway now. 

\section{Summary and conclusions} \label{sec:conclusions}

The objective of this study was to verify if the WB signal observed by IBEX comes up naturally in the simulations when a physically reasonable model perturbation of the interstellar plasma flow due to the heliosphere's presence is adopted and charge-exchange collisions between He atoms and \hepl atoms are assumed to operate. The positive outcome of this verification that we have obtained is another strong suggestion that WB is indeed the secondary population of ISN He, and that researching it may potentially bring important insight into the physical state of interstellar matter in front of the heliopause and into processes operating in this region.

To accomplish this objective, we have extended the Warsaw Test Particle Model (WTPM) to account for gains and losses of ISN He atoms due to charge exchange in a region extending far away beyond the heliopause, and we simulated the signal expected to be observed by IBEX-Lo for several different models of the plasma flow beyond the heliopause. The purpose was not to precisely reproduce the observed signal, but to study the sensitivity of the signal to selected aspects of the plasma flow past the heliopause. We adopted the observation conditions characteristic for four carefully chosen IBEX orbits (Figure~\ref{fig:orbitSelection}). In each of those orbits, a different aspect of the signal formation can be studied: pure WB, a mixture of WB and ISN He in comparable proportions, and orbits where ISN He dominates and WB is visible only in a subset of pixels. To add realism to our study, for numerical values of the density, bulk velocity, temperature, and ionization degree of the unperturbed gas we adopted those obtained from most up-to-date determinations, similarly as the direction and intensity of ISMF. 

The most important conclusion from our study is that a simulated WB signal builds up naturally as soon as one allows for charge-exchange collisions operating within the plasma ahead of the heliopause, where the plasma flow is disturbed by the heliosphere. When this disturbance is absent, only the ISN He component is present in the simulated signal (Figure~\ref{fig:hotModel}).

For an axially symmetric hydrodynamic model of the outer heliosheath, which predicts a relatively narrow sheath, the secondary population of ISN He would be visible for IBEX-Lo, but the simulated signal features important differences compared with the observations. Some of those differences, including an order of magnitude deficit in the pure-WB orbit, were partially removed by increasing the assumed size of the outer heliosheath. Other differences, including a strong deficit in the signal for the spin angle bins far away from the peak bins and the disturbed proportions between the signal peak heights among different orbits, remained (Figure~\ref{fig:fluxes-log2-Bonn}). 

For a MHD model with ISMF directed towards the IBEX Ribbon center and intensity in agreement with the current estimates, the resulting deformation of the heliosphere from axial symmetry removed the disagreement in the signal peak heights. Most of the remaining differences, especially in the wings of the signal, could be removed by increasing the plasma temperature inside the outer heliosheath. Furthermore, no up-scaling of the OHS size was required (Figure~\ref{fig:fluxes-log2-002sym}). This suggests that details of the temperature distribution of the plasma within OHS are more important for the creation of WB than the spatial extent of the OHS region.

These results suggest that even though the test models of the plasma flow we have used are not state of the art, apparently they are sufficiently realistic to grasp the most important features of the plasma flow beyond the heliopause. Details of the conditions in various regions of OHS can be studied in the future by detailed comparison of model results with observations. Due to the ballistics of ISN He atoms, the signal observed in individual pixels in the data from individual IBEX orbits can be mapped into distinct regions of OHS (Figure~\ref{fig:srcRegions}). Simultaneously, the portion of the observed signal corresponding mostly to the primary ISN He population does not seem to be morphologically biased by the secondary population and consequently the estimates of the temperature of the unperturbed local interstellar medium, obtained from analysis of direct sampling experiments, are not likely to be biased in an important way. 

Results of the paper suggest that the classical, comet-like model of the heliosphere, traveling through a $\sim 7500$~K local interstellar medium at $\sim 25$~\kms~with a distortion due to ISMF of $\sim 2 \mu$G directed towards the IBEX Ribbon center, is able to explain direct-sampling observations of interstellar neutral helium from IBEX.

{\em Acknowledgment} We thank Justyna Sok{\'o}{\l} for her help with the preparation of Figure~\ref{fig:srcRegions}. We are obliged to Romana Ratkiewicz for kindly providing us with an early version of the MHD code. This study was supported by Polish National Science Center grant 2015-19-B-ST9-01328.

\bibliographystyle{aasjournal}
\bibliography{iplbib}{}

\begin{thebibliography}{}
\expandafter\ifx\csname natexlab\endcsname\relax\def\natexlab#1{#1}\fi
\providecommand{\url}[1]{\href{#1}{#1}}

\bibitem[{Baranov \& Malama(1993)}]{baranov_malama:93}
Baranov, V.~B., \& Malama, Y.~G. 1993, \jgr, 98, 15157

\bibitem[{Barnett {et~al.}(1990)Barnett, Hunter, Kirkpatrick, Alvarez,
  Cisneros, \& Phaneuf}]{barnett_etal:90}
Barnett, C.~F., Hunter, H.~T., Kirkpatrick, M.~I., {et~al.} 1990, Atomic data
  for fusion. Collisions of {H}, {H}$_2$, {H}e and {L}i atoms and ions with
  atoms and molecules, Vol. ORNL-6086/V1 (Oak Ridge, Tenn.: Oak Ridge National
  Laboratories)

\bibitem[{Bertaux \& Blamont(1971)}]{bertaux_blamont:71}
Bertaux, J.~L., \& Blamont, J.~E. 1971, \aap, 11, 200

\bibitem[{{Burlaga} \& {Ness}(2014{\natexlab{a}})}]{burlaga_ness:14a}
{Burlaga}, L.~F., \& {Ness}, N.~F. 2014{\natexlab{a}}, \apjl, 795, L19

\bibitem[{{Burlaga} \& {Ness}(2014{\natexlab{b}})}]{burlaga_ness:14b}
---. 2014{\natexlab{b}}, \apj, 784, 146

\bibitem[{{Bzowski} {et~al.}(2014){Bzowski}, {Kubiak}, {H{\l}ond},
  {Sok{\'o}{\l}}, {Banaszkiewicz}, \& {Witte}}]{bzowski_etal:14a}
{Bzowski}, M., {Kubiak}, M.~A., {H{\l}ond}, M., {et~al.} 2014, \aap, 569, A8

\bibitem[{{Bzowski} {et~al.}(2008){Bzowski}, {M{\"o}bius}, {Tarnopolski},
  {Izmodenov}, \& {Gloeckler}}]{bzowski_etal:08a}
{Bzowski}, M., {M{\"o}bius}, E., {Tarnopolski}, S., {Izmodenov}, V., \&
  {Gloeckler}, G. 2008, \aap, 491, 7

\bibitem[{{Bzowski} {et~al.}(2009){Bzowski}, {M{\"o}bius}, {Tarnopolski},
  {Izmodenov}, \& {Gloeckler}}]{bzowski_etal:09a}
---. 2009, \ssr, 143, 177

\bibitem[{Bzowski {et~al.}(2013)Bzowski, Sok{\'o}{\l}, Kubiak, \&
  Kucharek}]{bzowski_etal:13b}
Bzowski, M., Sok{\'o}{\l}, J.~M., Kubiak, M.~A., \& Kucharek, H. 2013, \aap,
  557, A50

\bibitem[{Bzowski {et~al.}(2012)Bzowski, Kubiak, M{\"o}bius, Bochsler, Leonard,
  Heirtzler, Kucharek, Sok{\'{o}}{\l}, H{\l}ond, Crew, Schwadron, Fuselier, \&
  McComas}]{bzowski_etal:12a}
Bzowski, M., Kubiak, M.~A., M{\"o}bius, E., {et~al.} 2012, \apjs, 198, 12

\bibitem[{{Bzowski} {et~al.}(2015){Bzowski}, {Swaczyna}, {Kubiak},
  {Sok\'{o}{\l}}, {Fuselier}, {Galli}, {Heirtzler}, {Kucharek}, {Leonard},
  {McComas}, {M{\"o}bius}, {Schwadron}, \& {Wurz}}]{bzowski_etal:15a}
{Bzowski}, M., {Swaczyna}, P., {Kubiak}, M., {et~al.} 2015, \apjs, 220, 28

\bibitem[{{Chalov} {et~al.}(2010){Chalov}, {Alexashov}, {McComas}, {Izmodenov},
  {Malama}, \& {Schwadron}}]{chalov_etal:10a}
{Chalov}, S.~V., {Alexashov}, D.~B., {McComas}, D., {et~al.} 2010, \apjl, 716,
  L99

\bibitem[{{Czechowski} {et~al.}(2015){Czechowski}, {Grygorczuk}, \&
  {McComas}}]{czechowski_etal:15a}
{Czechowski}, A., {Grygorczuk}, J., \& {McComas}, D.~J. 2015, ArXiv e-prints,
  arXiv:1507.00540

\bibitem[{Fahr(1968)}]{fahr:68}
Fahr, H.~J. 1968, \apss, 2, 474

\bibitem[{Fahr(1978)}]{fahr:78}
---. 1978, \aap, 66, 103

\bibitem[{Fahr \& Bzowski(2004)}]{fahr_bzowski:04b}
Fahr, H.~J., \& Bzowski, M. 2004, \aap, 424, 263

\bibitem[{Fahr \& Mueller(1967)}]{fahr_mueller:67}
Fahr, H.~J., \& Mueller, K.~G. 1967, Z. Phys., 200, 343

\bibitem[{{Funsten} {et~al.}(2009){Funsten}, {Allegrini}, {Crew}, {DeMajistre},
  {Frisch}, {Fuselier}, {Gruntman}, {Janzen}, {McComas}, {M{\"o}bius},
  {Randol}, {Reisenfeld}, {Roelof}, \& {Schwadron}}]{funsten_etal:09b}
{Funsten}, H.~O., {Allegrini}, F., {Crew}, G.~B., {et~al.} 2009, Science, 326,
  964

\bibitem[{{Funsten} {et~al.}(2013){Funsten}, {DeMajistre}, {Frisch},
  {Heerikhuisen}, {Higdon}, {Janzen}, {Larsen}, {Livadiotis}, {McComas},
  {M{\"o}bius}, {Reese}, {Reisenfeld}, {Schwadron}, \&
  {Zirnstein}}]{funsten_etal:13a}
{Funsten}, H.~O., {DeMajistre}, R., {Frisch}, P.~C., {et~al.} 2013, \apj, 776,
  30

\bibitem[{{Funsten} {et~al.}(2015){Funsten}, {Bzowski}, {Cai}, {Dayeh},
  {DeMajistre}, {Frisch}, {Herrikhuisen}, {Hidgon}, {Janzen}, {Larsen},
  {Livadiotis}, {McComas}, {M{\"o}bius}, {Reese}, {Roelof}, {Reisenfeld},
  {Schwadron}, \& {Zirnstein}}]{funsten_etal:15a}
{Funsten}, H.~O., {Bzowski}, M., {Cai}, D.~M., {et~al.} 2015, \apj, 799, 68

\bibitem[{{Fuselier} {et~al.}(2009){Fuselier}, {Bochsler}, {Chornay}, {Clark},
  {Crew}, {Dunn}, {Ellis}, {Friedmann}, {Funsten}, {Ghielmetti}, {Googins},
  {Granoff}, {Hamilton}, {Hanley}, {Heirtzler}, {Hertzberg}, {Isaac}, {King},
  {Knauss}, {Kucharek}, {Kudirka}, {Livi}, {Lobell}, {Longworth}, {Mashburn},
  {McComas}, {M{\"o}bius}, {Moore}, {Moore}, {Nemanich}, {Nolin}, {O'Neal},
  {Piazza}, {Peterson}, {Pope}, {Rosmarynowski}, {Saul}, {Scherrer}, {Scheer},
  {Schlemm}, {Schwadron}, {Tillier}, {Turco}, {Tyler}, {Vosbury}, {Wieser},
  {Wurz}, \& {Zaffke}}]{fuselier_etal:09b}
{Fuselier}, S.~A., {Bochsler}, P., {Chornay}, D., {et~al.} 2009, \ssr, 146, 117

\bibitem[{{Galli} {et~al.}(2015){Galli}, {Wurz}, {Park}, {Kucharek},
  {M{\"o}bius}, {Schwadron}, {Sok\'{o}{\l}}, {Bzowski}, {Kubiak}, {Swaczyna},
  {Fuselier}, \& {McComas}}]{galli_etal:15a}
{Galli}, A., {Wurz}, P., {Park}, J., {et~al.} 2015, \apjs, 220, 30

\bibitem[{{Gloeckler} {et~al.}(2004){Gloeckler}, {M{\"o}bius}, {Geiss},
  {Bzowski}, {Chalov}, {Fahr}, {McMullin}, {Noda}, {Oka}, {Ruci{\'n}ski},
  {Skoug}, {Terasawa}, {von Steiger}, {Yamazaki}, \&
  {Zurbuchen}}]{gloeckler_etal:04b}
{Gloeckler}, G., {M{\"o}bius}, E., {Geiss}, J., {et~al.} 2004, \aap, 426, 845

\bibitem[{{Grygorczuk} {et~al.}(2014){Grygorczuk}, {Czechowski}, \&
  {Grzedzielski}}]{grygorczuk_etal:14a}
{Grygorczuk}, J., {Czechowski}, A., \& {Grzedzielski}, S. 2014, \apjl, 789, L43

\bibitem[{{Grzedzielski} {et~al.}(2010){Grzedzielski}, {Bzowski}, {Czechowski},
  {Funsten}, {McComas}, \& {Schwadron}}]{grzedzielski_etal:10b}
{Grzedzielski}, S., {Bzowski}, M., {Czechowski}, A., {et~al.} 2010, \apjl, 715,
  L84

\bibitem[{Gurnett {et~al.}(2013)Gurnett, Kurth, Burlaga, \&
  Ness}]{gurnett_etal:13a}
Gurnett, D.~A., Kurth, W.~S., Burlaga, L.~F., \& Ness, N.~F. 2013, Science,
  341, 1489

\bibitem[{{Heerikhuisen} {et~al.}(2015){Heerikhuisen}, {Zirnstein}, \&
  {Pogorelov}}]{heerikhuisen_etal:15a}
{Heerikhuisen}, J., {Zirnstein}, E., \& {Pogorelov}, N. 2015, \jgr, 120, 1516

\bibitem[{{Heerikhuisen} {et~al.}(2014){Heerikhuisen}, {Zirnstein}, {Funsten},
  {Pogorelov}, \& {Zank}}]{heerikhuisen_etal:14a}
{Heerikhuisen}, J., {Zirnstein}, E.~J., {Funsten}, H.~O., {Pogorelov}, N.~V.,
  \& {Zank}, G.~P. 2014, \apj, 784, 73

\bibitem[{{Heerikhuisen} {et~al.}(2010){Heerikhuisen}, {Pogorelov}, {Zank},
  {Crew}, {Frisch}, {Funsten}, {Janzen}, {McComas}, {Reisenfeld}, \&
  {Schwadron}}]{heerikhuisen_etal:10a}
{Heerikhuisen}, J., {Pogorelov}, N.~V., {Zank}, G.~P., {et~al.} 2010, \apjl,
  708, L126

\bibitem[{{Izmodenov} {et~al.}(2005){Izmodenov}, {Alexashov}, \&
  {Myasnikov}}]{izmodenov_etal:05a}
{Izmodenov}, V., {Alexashov}, D., \& {Myasnikov}, A. 2005, \aap, 437, L35

\bibitem[{{Izmodenov} {et~al.}(2003{\natexlab{a}}){Izmodenov}, {Gloeckler}, \&
  {Malama}}]{izmodenov_etal:03b}
{Izmodenov}, V., {Gloeckler}, G., \& {Malama}, Y. 2003{\natexlab{a}}, \grl, 30,
  3

\bibitem[{{Izmodenov} {et~al.}(2003{\natexlab{b}}){Izmodenov}, {Malama},
  {Gloeckler}, \& {Geiss}}]{izmodenov_etal:03a}
{Izmodenov}, V., {Malama}, Y.~G., {Gloeckler}, G., \& {Geiss}, J.
  2003{\natexlab{b}}, \apjl, 594, L59

\bibitem[{{Izmodenov} \& {Alexashov}(2006)}]{izmodenov_alexashov:06a}
{Izmodenov}, V.~V., \& {Alexashov}, D.~B. 2006, in AIP Conf. Proc. 858: Physics
  of the Inner Heliosheath, ed. J.~{Heerikhuisen}, V.~{Florinski}, G.~P.
  {Zank}, \& N.~V. {Pogorelov}, 14--19

\bibitem[{Izmodenov \& Alexashov(2015)}]{izmodenov_alexashov:15a}
Izmodenov, V.~V., \& Alexashov, D.~B. 2015, \apjs, 220, 32

\bibitem[{{Kausch} \& {Fahr}(1997)}]{kausch_etal:97a}
{Kausch}, T., \& {Fahr}, H.~J. 1997, \aap, 325, 828

\bibitem[{{Krimigis} {et~al.}(2009){Krimigis}, {Mitchell}, {Roelof}, {Hsieh},
  \& {McComas}}]{krimigis_etal:09a}
{Krimigis}, S.~M., {Mitchell}, D.~G., {Roelof}, E.~C., {Hsieh}, K.~C., \&
  {McComas}, D.~J. 2009, Science, 326, 971

\bibitem[{{Kubiak} {et~al.}(2014){Kubiak}, {Bzowski}, {Sok{\'o}{\l}},
  {Swaczyna}, {Grzedzielski}, {Alexashov}, {Izmodenov}, {Moebius}, {Leonard},
  {Fuselier}, {Wurz}, \& {McComas}}]{kubiak_etal:14a}
{Kubiak}, M.~A., {Bzowski}, M., {Sok{\'o}{\l}}, J.~M., {et~al.} 2014, \apjs,
  213, 29

\bibitem[{Kubiak {et~al.}(2016)Kubiak, Swaczyna, Bzowski, Sok{\'o}{\l},
  Fuselier, Galli, Heirtzler, Kucharek, Leonard, McComas, Park, Schwadron, \&
  Wurz}]{kubiak_etal:16a}
Kubiak, M.~A., Swaczyna, P., Bzowski, M., {et~al.} 2016, \apjs, 223, 35

\bibitem[{Lallement \& Bertaux(1990)}]{lallement_bertaux:90a}
Lallement, R., \& Bertaux, J.~L. 1990, \aap, 231, L3

\bibitem[{{Lallement} {et~al.}(1993){Lallement}, {Bertaux}, \&
  {Clarke}}]{lallement_etal:93a}
{Lallement}, R., {Bertaux}, J.-L., \& {Clarke}, J.~T. 1993, Science, 260, 1095

\bibitem[{{Lallement} {et~al.}(2005){Lallement}, {Qu{\' e}merais}, {Bertaux},
  {Ferron}, {Koutroumpa}, \& {Pellinen}}]{lallement_etal:05a}
{Lallement}, R., {Qu{\' e}merais}, E., {Bertaux}, J.~L., {et~al.} 2005,
  Science, 307, 1447

\bibitem[{{Lallement} {et~al.}(2010){Lallement}, {Qu{\'e}merais}, {Koutroumpa},
  {Bertaux}, {Ferron}, {Schmidt}, \& {Lamy}}]{lallement_etal:10a}
{Lallement}, R., {Qu{\'e}merais}, E., {Koutroumpa}, D., {et~al.} 2010, Twelfth
  International Solar Wind Conference, 1216, 555

\bibitem[{Leonard {et~al.}(2015)Leonard, M{\"o}bius, Bzowski, Fuselier,
  Heirtzler, Kubiak, Kucharek, Lee, McComas, Schwadron, \&
  Wurz}]{leonard_etal:15a}
Leonard, T.~W., M{\"o}bius, E., Bzowski, M., {et~al.} 2015, \apj, 804, 42

\bibitem[{{Lindsay} \& {Stebbings}(2005)}]{lindsay_stebbings:05a}
{Lindsay}, B.~G., \& {Stebbings}, R.~F. 2005, \jgr, 110, A12213

\bibitem[{{McComas} {et~al.}(2015{\natexlab{a}}){McComas}, {Bzowski}, {Frisch},
  {Galli}, {Izmodenov}, {Katushkina}, {Kubiak}, {Lee}, {Leonard}, {M{\"o}bius},
  {Park}, {Schwadron}, {Sok{\'o}{\l}}, {Swaczyna}, {Wood}, \&
  {Wurz}}]{mccomas_etal:15b}
{McComas}, D., {Bzowski}, M.~{Fuselier}, S., {Frisch}, P., {et~al.}
  2015{\natexlab{a}}, \apjs, 220, 22

\bibitem[{{McComas} {et~al.}(2015{\natexlab{b}}){McComas}, {Bzowski}, {Frisch},
  {Fuselier}, {Kubiak}, {Kucharek}, {Leonard}, {M{\"o}bius}, {Schwadron},
  {Sok{\'o}{\l}}, {Swaczyna}, \& {Witte}}]{mccomas_etal:15a}
{McComas}, D., {Bzowski}, M., {Frisch}, P., {et~al.} 2015{\natexlab{b}}, \apj,
  801, 28

\bibitem[{{McComas} {et~al.}(2009{\natexlab{a}}){McComas}, {Allegrini},
  {Bochsler}, {Bzowski}, {Collier}, {Fahr}, {Fichtner}, {Frisch}, {Funsten},
  {Fuselier}, {Gloeckler}, {Gruntman}, {Izmodenov}, {Knappenberger}, {Lee},
  {Livi}, {Mitchell}, {M{\"o}bius}, {Moore}, {Pope}, {Reisenfeld}, {Roelof},
  {Scherrer}, {Schwadron}, {Tyler}, {Wieser}, {Witte}, {Wurz}, \&
  {Zank}}]{mccomas_etal:09a}
{McComas}, D.~J., {Allegrini}, F., {Bochsler}, P., {et~al.} 2009{\natexlab{a}},
  \ssr, 146, 11

\bibitem[{{McComas} {et~al.}(2009{\natexlab{b}}){McComas}, {Allegrini},
  {Bochsler}, {Frisch}, {Funsten}, {Gruntman}, {Janzen}, {Kucharek},
  {M{\"o}bius}, {Reisenfeld}, \& {Schwadron}}]{mccomas_etal:09b}
---. 2009{\natexlab{b}}, \grl, 36, 12104

\bibitem[{{M{\"o}bius} {et~al.}(2013){M{\"o}bius}, {Liu}, {Funsten}, {Gary}, \&
  {Winske}}]{mobius_etal:13a}
{M{\"o}bius}, E., {Liu}, K., {Funsten}, H., {Gary}, S.~P., \& {Winske}, D.
  2013, \apj, 766, 129

\bibitem[{M{\"o}bius {et~al.}(2012)M{\"o}bius, Bochsler, Heirtzler, Kucharek,
  Lee, Leonard, Petersen, Schwadron, Valocvin, Wu, Bzowski, Kubiak, Fuselier,
  Saul, Wurz, McComas, \& Crew}]{mobius_etal:12a}
M{\"o}bius, E., Bochsler, P., Heirtzler, D., {et~al.} 2012, \apjs, 198, 11

\bibitem[{{M{\"o}bius} {et~al.}(2015){M{\"o}bius}, {Bzowski}, {Fuselier},
  {Heirtzler}, {Kubiak}, {Kucharek}, {Lee}, {Leonard}, {McComas}, {Schwadron},
  {Sok\'{o}{\l}}, \& {Wurz}}]{mobius_etal:15b}
{M{\"o}bius}, E., {Bzowski}, M., {Fuselier}, S.~A., {et~al.} 2015, \apjs, 220,
  24

\bibitem[{{M{\"u}ller} {et~al.}(2016){M{\"u}ller}, {M{\"o}bius}, \&
  {Wood}}]{mueller_etal:16a}
{M{\"u}ller}, H.-R., {M{\"o}bius}, E., \& {Wood}, B.~E. 2016, in Journal of
  Physics Conference Series, Vol. 767, Journal of Physics Conference Series,
  012019

\bibitem[{{M{\"u}ller} \& {Zank}(2003)}]{mueller_zank:03a}
{M{\"u}ller}, H.-R., \& {Zank}, G.~P. 2003, in American Institute of Physics
  Conference Series, Vol. 679, Solar Wind Ten, ed. {M.~Velli, R.~Bruno,
  F.~Malara, \& B.~Bucci}, 89--92

\bibitem[{{M{\"u}ller} \& {Zank}(2004)}]{mueller_zank:04a}
{M{\"u}ller}, H.-R., \& {Zank}, G.~P. 2004, \jgr, 109, A07104

\bibitem[{{Opher} {et~al.}(2015){Opher}, {Drake}, {Zieger}, \&
  {Gombosi}}]{opher_etal:15a}
{Opher}, M., {Drake}, J.~F., {Zieger}, B., \& {Gombosi}, T.~I. 2015, \apjl,
  800, L28

\bibitem[{Osterbart \& Fahr(1992)}]{osterbart_fahr:92}
Osterbart, R., \& Fahr, H.~J. 1992, \aap, 264, 260

\bibitem[{Park {et~al.}(2016)Park, Kucharek, M{\"o}bius, Galli, Kubiak,
  Bzowski, \& McComas}]{park_etal:16a}
Park, J., Kucharek, H., M{\"o}bius, E., {et~al.} 2016, \apj, 833, 130

\bibitem[{{Pogorelov} {et~al.}(2008){Pogorelov}, {Heerikhuisen}, \&
  {Zank}}]{pogorelov_etal:08a}
{Pogorelov}, N.~V., {Heerikhuisen}, J., \& {Zank}, G.~P. 2008, \apjl, 675, L41

\bibitem[{{Pogorelov} {et~al.}(2009){Pogorelov}, {Heerikhuisen}, {Zank}, \&
  {Borovikov}}]{pogorelov_etal:09b}
{Pogorelov}, N.~V., {Heerikhuisen}, J., {Zank}, G.~P., \& {Borovikov}, S.~N.
  2009, \ssr, 143, 31

\bibitem[{{Pogorelov} {et~al.}(2006){Pogorelov}, {Zank}, \&
  {Ogino}}]{pogorelov_etal:06b}
{Pogorelov}, N.~V., {Zank}, G.~P., \& {Ogino}, T. 2006, \apj, 644, 1299

\bibitem[{{Ratkiewicz} {et~al.}(1998){Ratkiewicz}, {Barnes}, {Molvik},
  {Spreiter}, {Stahara}, {Vinokur}, \& {Venkateswaran}}]{ratkiewicz_etal:98a}
{Ratkiewicz}, R., {Barnes}, A., {Molvik}, G.~A., {et~al.} 1998, \aap, 335, 363

\bibitem[{{Ratkiewicz} {et~al.}(2008){Ratkiewicz}, {Ben-Jaffel}, \&
  {Grygorczuk}}]{ratkiewicz_etal:08a}
{Ratkiewicz}, R., {Ben-Jaffel}, L., \& {Grygorczuk}, J. 2008, in Astronomical
  Society of the Pacific Conference Series, Vol. 385, Numerical Modeling of
  Space Plasma Flows, ed. N.~V. {Pogorelov}, E.~{Audit}, \& G.~P. {Zank},
  189--194

\bibitem[{Ripken \& Fahr(1983)}]{ripken_fahr:83a}
Ripken, H.~W., \& Fahr, H.~J. 1983, \aap, 122, 181

\bibitem[{Ruci{\'n}ski {et~al.}(2003)Ruci{\'n}ski, Bzowski, \&
  Fahr}]{rucinski_etal:03}
Ruci{\'n}ski, D., Bzowski, M., \& Fahr, H.~J. 2003, \ag, 21, 1315

\bibitem[{Saul {et~al.}(2012)Saul, Wurz, M{\"o}bius, Bzowski, Fuselier, Crew,
  Rodriguez, Leonard, McComas, Schwadron, Bochsler, \& Scheer}]{saul_etal:12a}
Saul, L., Wurz, P., M{\"o}bius, E., {et~al.} 2012, \apjs, 198, 14

\bibitem[{Saul {et~al.}(2013)Saul, Bzowski, Fuselier, Kubiak, McComas,
  M{\"o}bius, Sok{'o}{\l}, Rodr{\'i}guez, Scheer, \& Wurz}]{saul_etal:13a}
Saul, L., Bzowski, M., Fuselier, S., {et~al.} 2013, \apj, 767, 130

\bibitem[{{Schwadron} {et~al.}(2015){Schwadron}, {M{\"o}bius}, {Leonard},
  {Fuselier}, {Bzowski}, {Frisch}, {Heirtzler}, {Kubiak}, {Kucharek}, {Lee},
  {McComas}, {Rahmanifard}, {Sok\'{o}{\l}}, \& {Swaczyna}}]{schwadron_etal:15a}
{Schwadron}, N., {M{\"o}bius}, E., {Leonard}, T., {et~al.} 2015, \apjs, 220, 25

\bibitem[{{Schwadron} {et~al.}(2009){Schwadron}, {Bzowski}, {Crew}, {Gruntman},
  {Fahr}, {Fichtner}, {Frisch}, {Funsten}, {Fuselier}, {Heerikhuisen},
  {Izmodenov}, {Kucharek}, {Lee}, {Livadiotis}, {McComas}, {Moebius}, {Moore},
  {Mukherjee}, {Pogorelov}, {Prested}, {Reisenfeld}, {Roelof}, \&
  {Zank}}]{schwadron_etal:09b}
{Schwadron}, N.~A., {Bzowski}, M., {Crew}, G.~B., {et~al.} 2009, Science, 326,
  966

\bibitem[{{Schwadron} {et~al.}(2016){Schwadron}, {M{\"o}bius}, {McComas},
  {Bochsler}, {Bzowski}, {Fuselier}, {Livadiotis}, {Frisch}, {M{\"u}ller},
  {Heirtzler}, {Kucharek}, \& {Lee}}]{schwadron_etal:16a}
{Schwadron}, N.~A., {M{\"o}bius}, E., {McComas}, D.~J., {et~al.} 2016, \apj,
  828, 81

\bibitem[{{Slavin} \& {Frisch}(2008)}]{slavin_frisch:08a}
{Slavin}, J.~D., \& {Frisch}, P.~C. 2008, \aap, 491, 53

\bibitem[{{Sok{\'o}{\l}} \& {Bzowski}(2014)}]{sokol_bzowski:14a}
{Sok{\'o}{\l}}, J.~M., \& {Bzowski}, M. 2014, ArXiv e-prints, arXiv:1411.4826

\bibitem[{{Sok\'{o}{\l}} {et~al.}(2016){Sok\'{o}{\l}}, {Bzowski}, {Kubiak}, \&
  {M{\"o}bius}}]{sokol_etal:16a}
{Sok\'{o}{\l}}, J.~M., {Bzowski}, M., {Kubiak}, M., \& {M{\"o}bius}, E. 2016,
  \mnras, 458, 3691

\bibitem[{{Sok\'{o}{\l}} {et~al.}(2015{\natexlab{a}}){Sok\'{o}{\l}}, {Kubiak},
  {Bzowski}, \& {Swaczyna}}]{sokol_etal:15b}
{Sok\'{o}{\l}}, J.~M., {Kubiak}, M., {Bzowski}, M., \& {Swaczyna}, P.
  2015{\natexlab{a}}, \apjs, 220, 27

\bibitem[{{Sok\'{o}{\l}} {et~al.}(2015{\natexlab{b}}){Sok\'{o}{\l}},
  {Swaczyna}, {Bzowski}, \& {Tokumaru}}]{sokol_etal:15d}
{Sok\'{o}{\l}}, J.~M., {Swaczyna}, P., {Bzowski}, M., \& {Tokumaru}, M.
  2015{\natexlab{b}}, \solphys, 290, 2589

\bibitem[{{Stone} {et~al.}(2013){Stone}, {Cummings}, {McDonald}, {Heikkila},
  {Lal}, \& {Webber}}]{stone_etal:13a}
{Stone}, E.~C., {Cummings}, A.~C., {McDonald}, F.~B., {et~al.} 2013, Science,
  341, 150

\bibitem[{{Swaczyna} {et~al.}(2016{\natexlab{a}}){Swaczyna}, {Bzowski},
  {Christian}, {Funsten}, {McComas}, \& {Schwadron}}]{swaczyna_etal:16a}
{Swaczyna}, P., {Bzowski}, M., {Christian}, E.~R., {et~al.} 2016{\natexlab{a}},
  \apj, 823, 119

\bibitem[{{Swaczyna} {et~al.}(2016{\natexlab{b}}){Swaczyna}, {Bzowski}, \&
  {Sok{\'o}{\l}}}]{swaczyna_etal:16b}
{Swaczyna}, P., {Bzowski}, M., \& {Sok{\'o}{\l}}, J.~M. 2016{\natexlab{b}},
  \apj, 827, 71

\bibitem[{{Swaczyna} {et~al.}(2015){Swaczyna}, {Bzowski}, {Kubiak},
  {Sok\'{o}{\l}}, {M{\"o}bius}, {Leonard}, {Heirtzler}, {Kucharek},
  {Schwadron}, {Fuselier}, \& {McComas}}]{swaczyna_etal:15a}
{Swaczyna}, P., {Bzowski}, M., {Kubiak}, M., {et~al.} 2015, \apjs, 220, 26

\bibitem[{{Witte}(2004)}]{witte:04}
{Witte}, M. 2004, \aap, 426, 835

\bibitem[{{Wood} {et~al.}(2015){Wood}, {M{\"u}ller}, \&
  {Witte}}]{wood_etal:15a}
{Wood}, B.~E., {M{\"u}ller}, H.-R., \& {Witte}, M. 2015, \apj, 801, 62

\bibitem[{Wu \& Judge(1979)}]{wu_judge:79a}
Wu, F.~M., \& Judge, D.~L. 1979, \apj, 231, 594

\bibitem[{{Zank}(2015)}]{zank:15a}
{Zank}, G.~P. 2015, \araa, 53, 449

\bibitem[{Zank {et~al.}(2013)Zank, Heerikhuisen, Wood, Pogorelov, Zirnstein, \&
  McComas}]{zank_etal:13a}
Zank, G.~P., Heerikhuisen, J., Wood, B.~E., {et~al.} 2013, \apj, 763, 20

\bibitem[{{Zank} {et~al.}(2014){Zank}, {Hunana}, {Mostafavi}, \&
  {Goldstein}}]{zank_etal:14b}
{Zank}, G.~P., {Hunana}, P., {Mostafavi}, P., \& {Goldstein}, M.~L. 2014, \apj,
  797, 87

\bibitem[{Zank {et~al.}(1996)Zank, Pauls, Williams, \& Hall}]{zank_etal:96a}
Zank, G.~P., Pauls, H.~L., Williams, L.~L., \& Hall, D. 1996, \jgr, 101, 21636

\bibitem[{{Zirnstein} {et~al.}(2016){Zirnstein}, {Heerikhuisen}, {Funsten},
  {Livadiotis}, {McComas}, \& {Pogorelov}}]{zirnstein_etal:16b}
{Zirnstein}, E.~J., {Heerikhuisen}, J., {Funsten}, H.~O., {et~al.} 2016, \apjl,
  818, L18

\end{thebibliography}


\end{document}